\documentclass[fleqn,usenatbib]{rasti}

\usepackage{newtxtext,newtxmath}

\usepackage[T1]{fontenc}

\DeclareRobustCommand{\VAN}[3]{#2}
\let\VANthebibliography\thebibliography
\def\thebibliography{\DeclareRobustCommand{\VAN}[3]{##3}\VANthebibliography}

\usepackage{graphicx}	
\usepackage{amsmath}	
\usepackage{booktabs}

\newcommand{\thethree}{{\sc The300}}

\newcommand{\gadgetx}{\textsc{Gadget-X}}

\newcommand{\gizmo}{\textsc{Gizmo}}
\newcommand{\gadget}{\textsc{Gadget}}
\newcommand{\music}{\textsc{Music}}

\title[DL generated observations from simulations]{Deep Learning generated observations of galaxy clusters from dark-matter-only simulations}

\author[A. Caro et al.]{
Andrés Caro$^{1,2}$\thanks{andresf.caro@estudiante.uam.es},
Daniel de Andres$^{1,2}$\thanks{daniel.deandres@uam.es},
Weiguang Cui$^{1,2,3}$\thanks{Talento-CM fellow}, 
Gustavo Yepes$^{1,2}$, 
Marco De Petris$^{4}$,
\newauthor
Antonio Ferragamo$^{4,5}$,
Félicien Schiltz$^{6}$
and Amélie Nef$^{6}$
\\
$^{1}$ Departamento de Física Teórica, M-8, Universidad Autónoma de Madrid, Cantoblanco 28049, Madrid, Spain\\
$^{2}$ Centro de Investigación Avanzada en Física Fundamental,(CIAFF), Universidad Aut\'{o}noma de Madrid, Cantoblanco, 28049 Madrid, Spain\\
$^{3}$ Institute for Astronomy, University of Edinburgh, Royal Observatory, Edinburgh EH9 3HJ, UK \\
$^{4}$ Dipartimento di Fisica, Sapienza Universitá di Roma, Piazzale Aldo Moro, 5-00185 Roma, Italy \\
$^{5}$ Dipartimento di Fisica ‘E. Pancini’, Università degli Studi di Napoli Federico II, Via Cinthia, 21, I-80126 Napoli, Italy\\
$^{6}$ EURANOVA, Mont-Saint-Guibert, Belgium\\
}

\date{Accepted XXX. Received YYY; in original form ZZZ}

\pubyear{2024}

\begin{document}
\label{firstpage}
\pagerange{\pageref{firstpage}--\pageref{lastpage}}
\maketitle

\begin{abstract}
Hydrodynamical simulations play a fundamental role in modern cosmological research, serving as a crucial bridge between theoretical predictions and observational data. However, due to their computational intensity, these simulations are currently constrained to relatively small volumes. Therefore, this study investigates the feasibility of utilising dark matter-only simulations to generate observable maps of galaxy clusters using a deep learning approach based on the U-Net architecture. We focus on reconstructing Compton-y parameter maps (SZ maps) and bolometric X-ray surface brightness maps (X-ray maps) from total mass density maps. We leverage data from \textsc{The Three Hundred} simulations, selecting galaxy clusters ranging in mass from $10^{13.5} h^{-1}M_{\odot}\leq M_{200} \leq 10^{15.5}  h^{-1}M_{\odot}$.
Despite the machine learning models being independent of baryonic matter assumptions, a notable limitation is their dependency on the underlying physics of hydrodynamical simulations. To evaluate the reliability of our generated observable maps, we employ various metrics and compare the observable-mass scaling relations. For clusters with masses greater than $2 \times 10^{14} h^{-1} M_{\odot}$, the predictions show excellent agreement with the ground-truth datasets, with percentage errors averaging (0.5±0.1)\% for the parameters of the scaling laws. 


\end{abstract}




\section{Introduction}\label{sec-1}

Galaxy clusters are the largest gravitationally bound structures in the universe. 
The formation of these clusters involves the collapse of the most significant overdensities within the initial density field, driven by intricate gravitational dynamics and baryonic matter interactions associated with galaxy formation \citep[see, e.g.,][]{Allen2011, Kravtsov2012}.
Galaxy clusters provide valuable insights into several key aspects, including the growth of cosmic structure \citep{Walker2019}, the determination of cosmological parameters \citep{PlanckCollaboration2016, Salvati2022, Chiu2023}, and the evolution of the hot intracluster medium \citep[ICM][]{Jones1990, Bohringer2010}.
They also provide insights into the oldest stellar populations \citep{Soares2019} and the impact of supermassive black holes on both galaxies and surrounding gas \citep{Ferrarese2000}.
Thus, extensive research into galaxy clusters has greatly enhanced our understanding of cosmology and dark matter \citep{Battistelli2016}.

Observationally, galaxy clusters can be probed using various multi-wavelength observations. For instance,
 The Sloan Digital Sky Survey (SDSS\footnote{\url{https://www.sdss.org}}), the Hubble Space Telescope (HST\footnote{\url{https://science.nasa.gov/mission/hubble}}) and James Webb Space Telescope (JWST\footnote{\url{https://www.jwst.nasa.gov}}) significantly contribute to our understanding of the optical spectrum.
 Instruments such as XMM-Newton \citep{CHEX-MATECollaboration2021} and eROSITA \citep{Liu2022} explore the ICM through X-ray observations.  The Planck spacecraft \citep{PlanckCollaboration2016b}, the South Pole Telescope \citep{Bleem2020} and the Atacama Cosmology Telescope \citep[ACT,][]{ACT} offer insights into microwave frequencies via the Sunyaev-Zel'dovich \citep[SZ,][]{SZeffect} effect.

In cosmological research, bridging the gap between theoretical predictions and empirical data by populating dark matter-only simulations with observational baryonic properties has become a fundamental endeavour.
Various computational techniques, such as Halo Occupation Distribution models \citep[HOD,][]{Peacock2000, Kravtsov2004}, Semi-Analytic Models \citep[SAMs,][]{Croton2007, Benson2012}, and Subhalo Abundance Matching \citep[SHAM,][]{Vale2004, Conroy2006}, have been employed to address this challenge.
These methods establish a connection between the masses of dark matter halos and the properties of the baryonic matter they contain \citep{Schneider2019}.
While these methods often align with observations, they frequently fail to capture the complexities of galaxy formation and evolution.

Cosmological hydrodynamical simulations have emerged as invaluable tools to address some of these limitations.
These simulations span a broad range of scales and track numerous physical processes \citep{Borgani2011, Vogelsberger2020}.
They are crucial for calibrating advanced mass estimation algorithms \citep{deAndres2022}, investigating the correlation between morphology and dynamical state within clusters \citep{DeLuca2021}, tracking the evolutionary trajectory of kinematic properties in stellar components \citep{Mostoghiu2021}, and exploring the intricate relationship between star formation and various physical processes in galaxy clusters \citep{Hough2023}.
However, conducting full-box hydrodynamical simulations for galaxy clusters presents significant challenges, primarily due to the high processing cost associated with modelling the complex fluid dynamics from large cosmic scales to individual galaxies.
This requires extensive use of high-performance computing resources and substantial financial investment.
Additionally, resolution constraints and numerical instability pose significant barriers, particularly when modelling large-scale and small-scale structures simultaneously.
Incorporating phenomena such as gas clumping \citep{Nagai2011}, AGN feedback \citep{Gitti2012}, and star formation \citep{Bassini2020} further complicates the process, necessitating analytical prescriptions known as “sub-grid” physics.

In recent years, machine learning (ML) techniques have become instrumental in revealing complex patterns and extracting intricate correlations across various scientific fields, including astronomy and cosmology \citep{Smith2023}.
These applications include transient type identification \citep{Goldstein2015}, photometric redshift estimation \citep{Pasquet2019}, source categorization during reionization \citep{Hassan2019}, lensing signal detection in images \citep{Lanusse2018}, and galaxy morphology classification \citep{Huertas-Company2015, Dieleman2015, DominguezSanchez2018}.
Furthermore, ML techniques have been employed in regression tasks such as refining cluster mass infference \citep{Ntampaka2016, Ho2019, Yan2020, deAndres2022, Ferragamo2023}, determining the duration of reionization \citep{LaPlante2019}, and improving cosmological parameter constraints through noiseless catalogues \citep{Gupta2018}.

ML has also proven to be valuable for generating synthetic observations, effectively emulating results from computationally intensive simulations.
Neural networks, for example, have been utilized to understand the complex relationship between dark matter halos and galaxy properties, leading to the creation of extensive idealized catalogues from a limited number of high-fidelity simulations \citep[e.g.,][]{Berger2019, Bernardini2022, Lovell2022, deAndres2023}.
Additionally, generative models such as Generative Adversarial Networks (GANs) and Variational Autoencoders (VAEs) have been employed to produce synthetic observations that closely resemble real data \citep[e.g.,][]{Mustafa2019, Rothschild2022}.
Hybrid approaches combining ML techniques with traditional methods have shown potential in enhancing both accuracy and efficiency \citep{Xu2013, Hearin2020, Moews2021}.

The primary objective of this work is to evaluate the effectiveness of using dark matter-only simulations to predict observable maps of galaxy clusters.
To accomplish this task, we specifically utilize U-Net convolutional neural networks, a widely used deep learning architecture.
Our approach focuses on reconstructing Compton-y parameter maps and bolometric X-ray surface brightness maps from dark matter simulations, with an emphasis on map-to-map predictions.
In this context, we aim to advance our understanding of the capabilities and limitations of deep learning techniques in mapping the observable properties of galaxy clusters.

The paper is outlined as follows: in the following Section \ref{sec-2}, we describe our hydrodynamical and dark-matter-only simulations and how the training and test datasets are generated. Then, in Section \ref{sec-3}, we describe the characteristics of the considered deep learning algorithm and the training procedure.  In Section \ref{sec-4}, we show the main results of this work, covering an analysis of the generated observations from DM-only simulations and the performance of the ML models with various metrics and scaling relations, as well as showing a graphical interpretation. Finally, in Section \ref{sec-5}, we discuss the implications of our findings.


\section{Dataset}\label{sec-2}

\begin{table}
\centering
\begin{tabular}{lll}
\hline
Signal & Label & Units  \\ \hline
Total mass density (DM+baryons) & \texttt{total\_mass} & $h^{-1}$ $M_{\odot}$ kpc$^{-2}$   \\
X-ray Bolometric & \texttt{Xray} & erg s$^{-1}$ kpc$^{-2}$  \\
 Compton-$y$ Parameter & \texttt{SZ} & \textit{dimensionless}  \\ \hline
\end{tabular}%
\caption{Dataset units and the labels used
throughout this paper.}
\label{tab:signals} 
\end{table}

\subsection{The Three Hundred Simulations}\label{sec-2.1}

This study uses \textsc{The Three Hundred} (\thethree) simulations, which consists of 324 full-physics hydrodynamical ``zoom-in'' re-simulations of galaxy clusters \citep{Cui2018, Cui2022}. The unique setups of \thethree{}, such as the extensive surrounding area of $15\, h^{-1} \text{Mpc}$ around the central cluster, allow for the study of the filamentary structures connecting to the cluster \citep{Kuchner2020, Kuchner2021, Rost2021, Rost2024, Santoni2024}, the large sample of clusters facilitates statistical studies of backsplash galaxies \citep{Arthur2019, Haggar2020, Knebe2020}, lensing studies \citep{Vega-Ferrero2021, Herbonnet2022, EuclidCollaboration2024, Miren2024}, cluster dynamical state \citep{Capalbo2021,  DeLuca2021, Li2022, Zhang2022,2024Haggar}, cluster profiles \citep{Mostoghiu2019, Li2020, Baxter2021}, cluster mass \citep{Li2021, Gianfagna2023}. In addition, {\sc The300} simulations have been extensively used for training different ML models 
\citep{deAndres2022,deAndres2023,Contreras2023ML,Ferragamo2023,Iqbal2023,Arendt2024,deAndres2024}. These studies show  that the galaxy clusters in our simulations form a very rich dataset specially for training ML models.

\thethree{} is derived from the DM-only MultiDark Simulation \citep[MDPL2,][]{Klypin2016}. It consists of a cube of volume $(1 h^{-1} \text{Gpc})^3$ with $3840^3$ dark matter particles, each having a mass of $1.5 \times 10^9 h^{-1} M_\odot$. The simulation adopts cosmological parameters from the Planck 2015 results \citep{PlanckCollaboration2016}. \thethree{} zoom-in re-simulations are centred at the 324 most massive halos of the MDPL2 simulation at $z=0$, where all particles within a sphere of radius $15\, h^{-1} \text{Mpc}$ from the cluster centre are traced back to their initial conditions. These particles are divided into dark matter and gas particles based on the assumed cosmic baryon fraction. Additionally, low-resolution dark-matter particles with variable masses simulate the effects of the large-scale structure at greater distances beyond this sphere of interest. The final simulation dataset includes re-simulated cluster regions using two codes: the SPH (Smoothed Particle Hydrodynamics) code \textsc{Gadget-X} \citep{Rasia2015, Beck2016} and the MFM (Meshless Finite Mass) code \textsc{Gizmo-Simba} \citep{Hopkins2015, Dave2019}. A total of 129 snapshots are stored from each run, from redshift $z=0$ up to $z=17$. For a more detailed understanding of \textsc{Gadget-X} and \textsc{Gizmo-Simba} runs (\gadget{}  and \gizmo{} for simplicity), as well as the distinctions between them, we recommend the work of \cite{Cui2022}.

We employ the same dataset selection as described in \cite{deAndres2024}, where the criteria include selecting halos with a mass  $M_{200}>10^{13.5} h^{-1} M_\odot$ at a redshift close to zero. These halos are identified using the AMIGA Halo Finder package \citep{Knollmann2009}, where a contamination indicator of \texttt{fMhires} $ \simeq 1$ is used for ensuring the absence of heavy or low resolution dark matter particles infiltrating from outside the re-simulated volume.
Initially, the halos are selected within the \gadget{} dataset, ensuring that their counterparts in \gizmo{} match both quantity and mass distribution.
Moreover, to improve the statistical robustness of underrepresented clusters, the data selection includes objects from additional snapshots at various redshifts ($z = 0.022, 0.045, 0.069, 0.093$, and $0.117$), thus ensuring a uniform distribution of mass. To study the dependence of our model on mass, the resulting dataset is roughly segmented into three equally sized bins, delineated by two mass demarcations $M_{200} = 10^{14.05} h^{-1} M_\odot$ and $M_{200} = 10^{14.65} h^{-1} M_\odot$. Here, $M_{200}$ refers to the mass inside a sphere whose density is 200 times the universe's critical density at the corresponding redshift. After this selection process, we have 2518 halos selected from \gadget{} and 2523 from \gizmo{}, preserving an uniform distribution of mass across both datasets. Additionally, we created a new combined dataset from both simulations which we will refer to as \gizmo{}+\gadget{}.

\subsection{Simulated multiview images}
\label{sec-2.2}

After selecting the halos, we consider their total mass and two distinct simulated images: Compton-y parameter maps, given by the SZ effect (in short SZ maps), and bolometric X-ray surface brightness maps (in short X-ray maps). These two distinct maps represent an idealized theoretical observation of the gas component of clusters of galaxies. Table \ref{tab:signals} shows the units and short labels of the signal maps. These simulated maps have also been considered in our previous works for the tasks of inferring mass profiles from SZ maps \citep{Ferragamo2023},  mapping the matter distribution \citep{deAndres2024} and identifying galaxy cluster mergers \citep{Arendt2024}. Although the training dataset is the same as it was considered in \cite{deAndres2024}, a very brief summary is provided in the following lines:

\begin{itemize}

     \item The generation of \textbf{total mass density maps} is carried out by projecting the sum of the masses of all particles, that is, gas, star, dark matter, and black hole particles in the observer’s line of sight.

    \item The \textbf{SZ maps} are computed by using the publicly available library {\sc PYMSZ}\footnote{\url{https://github.com/weiguangcui/pymsz}} \citep{Cui2018}, that integrates the pressure field along the line of sight, i.e.:
    \begin{equation}
        y = \frac{\sigma_{\text{T}}k_{\text{B}}}{m_{\text{e}}c^{2}}\int n_{\text{e}}T_{\text{e}}dl \text{ .}
        \label{y-SZ}
    \end{equation}
    here $\sigma_{\text{T}}$ is the Thomson cross section, $k_{\text{B}}$ is the Boltzmann constant, $c$ the speed of light, $m_{\text{e}}$ the electron rest-mass, $n_{\text{e}}$ the electron number density and $T_{\text{e}}$ the electron temperature.
    
     
     \item The \textbf{X-ray maps} represent the radiative emission of the Intracluster Medium (ICM). Using a nearest grid point technique, the simulated particles within each cluster halo are projected onto a 2D grid to create these maps. Assuming an ideal monatomic gas with an isentropic expansion factor $\gamma = {5}/{3}$, gas temperatures are obtained from the hydrodynamic simulations using the internal energies and gas masses. Power curves are created using the XrayLum\footnote{ \url{https://github.com/rennehan/xraylum}} Python module that is builded upon the PyAtomDB\footnote{ \url{https://atomdb.readthedocs.io/en/master/}} package \citep{Foster2020}. This program computes, under the assumption of a hydrogen fraction of 0.76, the cooling curves for electron-electron bremsstrahlung and the emissions from H, He, C, N, Si, and Fe within the energy range of $0.5$ to $10$ keV. The individual power contributions at each gas location are calculated using the temperature with the following equation:

    \begin{equation}
        P_{Z}(kT) = \sum_{i=0}^{1000} S_{i}(k_{B}T) \cdot E_{\text{avg},i},
        \label{powerSpec}
    \end{equation}
    
    where $S_{i}(k_{B}T)$ is the spectrum at energy $i$ for a given energy with temperature $T$, and $E_{\text{avg},i}$ is the average energy from energy $i$ to $i+1$. 
    
    These contributions are linearly combined for each element and multiplied by a gas factor ($n_\text{H} n_\text{e} V$), representing the number of particles at that location, as follows:

    \begin{equation}
         \begin{split}
        L_{X,\text{bol}} = & \left[ \sum_{Y \in \{\text{e-e, H, He}\}} P_{Y}(kT) \right. \quad + \\
        & \left. \sum_{Z \in \{\text{C, N, Si, Fe}\}} A_Z \cdot P_{Z}(kT) \right] n_\text{H} n_\text{e} V,
        \end{split}
    \label{Xray-Lum}
    \end{equation}

    where $A_{Z}$ are the abundances of the elements $Z$, $n_\text{H}$ is the hydrogen number density, $n_\text{e}$ is the electron number density, $V$ is the volume of the gas particles, and the term "e-e" refers to electron-electron bremsstrahlung. The total power is integrated along the line of sight and binned into an image, with the pixel count representing the X-ray luminosity.

\end{itemize}

\begin{figure}
\centering
\includegraphics[width=1\columnwidth]{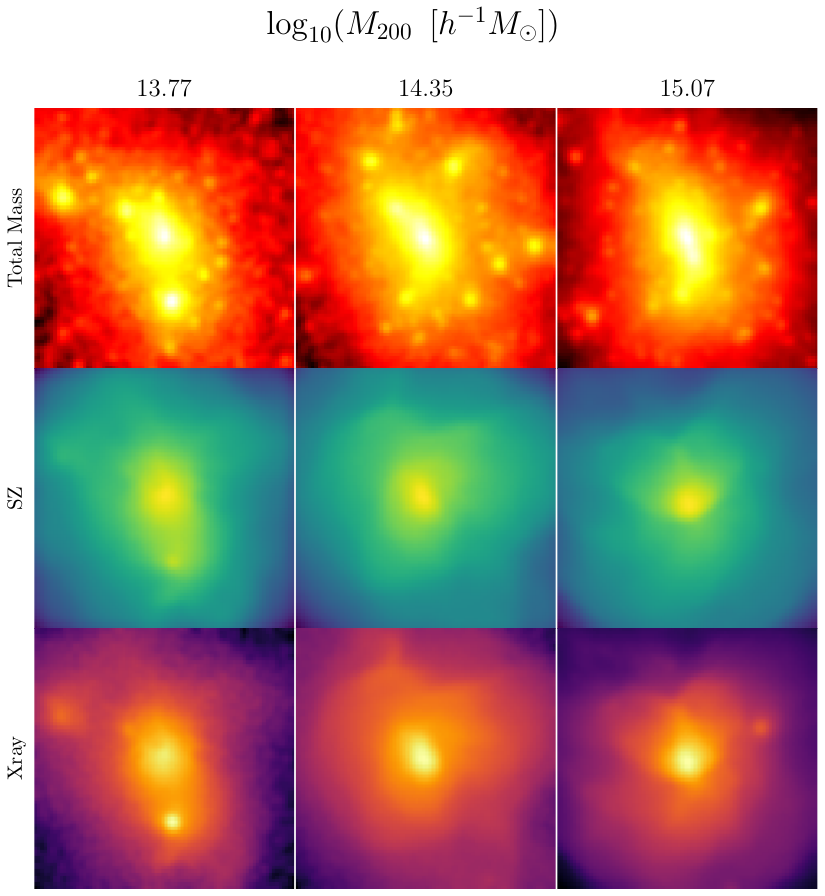}
\caption{Examples of maps from the datasets employed to
train the U-Net model. The first row depicts the total mass maps, while the last two show the simulated images of the Sunyaev-Zel'dovich (SZ) effect and bolometric X-ray emissions, respectively. Each map in the dataset spans a size of $2 \times R_{200}$. The mass $M_{200}$ of each cluster is indicated in the top label. The maps are shown in log$_{10}$ scale to better highlight the spatial and structural details.}
\label{fig:DataSet_Multiview} 
\end{figure}

The initial maps were generated with dimensions of $2R_{200}\times 2R_ {200} =640 \times 640$ pixels and subsequently smoothed using a Gaussian kernel with a full width at half maximum (FWHM) of $0.015 \times R_{200}$. Following the smoothing process, all maps were downsampled to a reduced resolution of $80 \times 80$ pixels. It is important to note that these maps are derived from particles confined within a cubic volume of length $2\times R_{200}$, resulting in the dimensions of our final $80 \times 80$ pixel images being $2 R_{200} \times 2R_{200}$. 
We show examples of our simulated maps for different cluster masses in Figure \ref{fig:DataSet_Multiview}. The objective is to apply a deep learning model to total mass maps to infer SZ or X-ray maps, as illustrated from top to bottom in the figure. This approach aims to capture and represent the non-linear relationships between the input (total mass maps) and the output (SZ or X-ray maps). The deep learning model allows for the straightforward conversion of a single input mass map into a corresponding observable map. Section \ref{sec-3} provides detailed information on the model and training process.

\begin{figure*}
\centering
\includegraphics[width=1\textwidth]{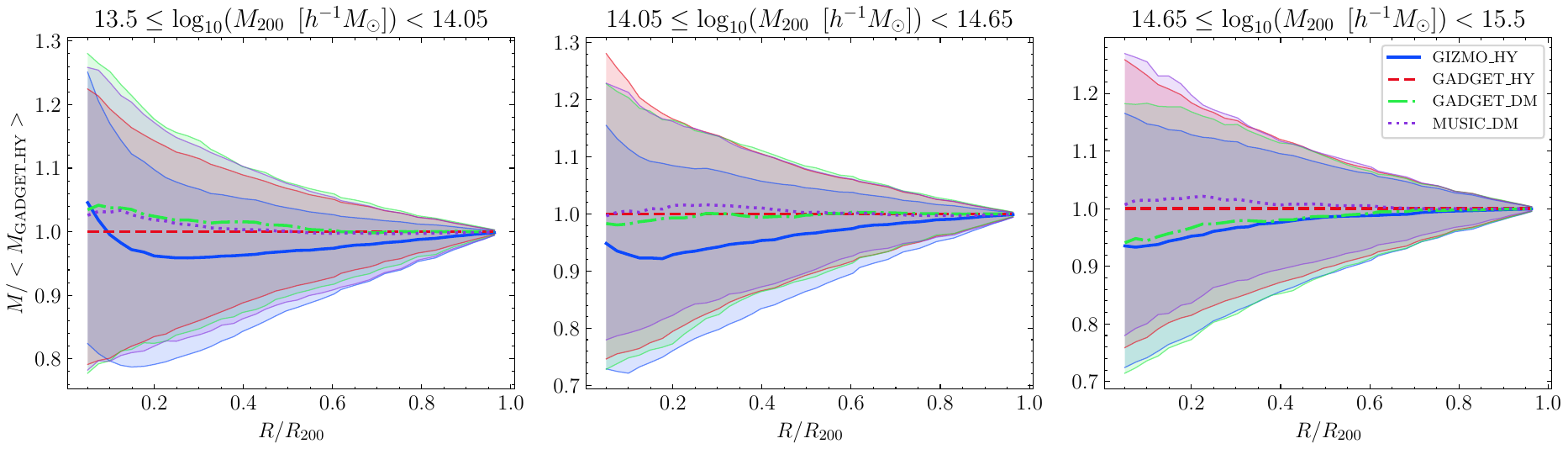}
\caption{ Total mass profiles are depicted for four distinct simulations: two hydrodynamical simulations (\textsc{Gizmo\_hy} and \textsc{Gadget\_hy}) and two N-body dark matter-only simulations (\textsc{Gadget\_dm} and \textsc{Music\_dm}), outlined in Section \ref{sec-2}. The y-axis denotes mass normalized by the median radial mass of the \textsc{Gadget\_hy} clusters, while the x-axis represents the cluster radius normalized by their corresponding $R_{200}$. Each graph, from left to right, corresponds to the three intervals specified in the plot titles. The lines illustrate the median values per each mass range, while the shaded regions denote from $16^{\text{th}}$ to $84^{\text{th}}$ percentiles.}
\label{fig:4sim_density_profiles} 
\end{figure*}

\subsection{N-body dark-matter-only simulations} \label{sec-2.3}

To evaluate our model on dark matter-only simulations, we specifically employ the DM-only variant of \thethree{} simulation, which uses the \gadgetx{} code. This code supports collisionless simulations in addition to smoothed particle hydrodynamics. This DM-only version of \thethree{} simulation (referred to as "\gadget{}-X DM-only" from now on) features a dark matter particle resolution of $1.5 \times 10^9 h^{-1} M_\odot$. We use the identical data selection process described in Section \ref{sec-2.1} to filter the clusters, thus having a total of 2155 halos.

\subsection{MUSIC simulation}

Furthermore, we also incorporate data from the Marenostrum-MultiDark Simulations of Galaxy Clusters (\music{}). Specifically dark matter-only particles of the re-simulations within the MultiDark Database (\music{} DM-only), which was conducted using the Adaptive Refinement Tree (ART) technique. 
This simulation, being hydrodynamic with multiple versions, has re-simulated clusters with non-radiative (NR) physics (i.e., without cooling, star formation, or supernovae feedbacks), where only SPH and gravity forces apply. For this study, we select only the positions of the dark matter particles. We rescale the masses by summing the original masses $m_{\text{DM}} = 9.01 \times 10^8 h^{-1} M_\odot$ and $m_{\text{SPH}} = 1.9 \times 10^8 h^{-1} M_\odot$, assigning them to the dark matter particles, resulting in a new mass of $m_{\text{DM}} = 1.091 \times 10^9 h^{-1} M_\odot$. This simulation features a low-resolution version ($256^3$ particles) of the MultiDark simulation and adopts the cosmological parameters from 7-years release of WMAP \citep{WMAP2011}, which slightly differ from those used in \thethree{} project. We perform a similar selection process as described for the primary dataset in Section \ref{sec-2.1}, focusing on the snapshot of the re-simulations at redshift $z=0$. This selection yields 673 halos with three distinct lines of sight. For a complete understanding of the \music{} simulations, we refer the reader to the work of \cite{Sembolini2013}. In this work, this simulation is considered dark-matter-only since the astrophysical effects implemented are minimal and it serves as and independent test set for our models.

Since these simulations focus exclusively on the gravitational effects of dark matter particles, the resulting mass distributions are less complex compared to those in hydrodynamical simulations. For instance, the \gadget{} simulation is notably successful in reproducing observed gas properties and relations, whereas the \gizmo{} simulation excels in replicating galaxy stellar properties, as noted by \cite{Cui2022}. Consequently, we expect differences in the mass maps depending on the simulation used. This distinction is evident in Figure \ref{fig:4sim_density_profiles}, which illustrates the mass radial profiles of galaxy clusters from various simulations. The figure compares the mass distribution profiles from hydrodynamical and dark matter-only simulations across three different mass ranges: $13.5 \leq \log_{10}(M_{200} [h^{-1}M_\odot]) < 14.05$, $14.05 \leq \log_{10}(M_{200} [h^{-1}M_\odot]) < 14.65$, and $14.65 \leq \log_{10}(M_{200} [h^{-1}M_\odot]) < 15.5$. The profiles are normalized by the median mass profile of the \gadget{} hydrodynamical simulation. The hydrodynamical simulations (\textsc{Gizmo\_hy} and \textsc{Gadget\_hy}) and dark matter-only simulations (\textsc{Gadget\_dm} and \textsc{Music\_dm}) show variations in mass distribution, reflecting the influence of baryonic physics in the hydrodynamical models. Notably, the differences in the mass profiles also depend on the total mass of the clusters, as indicated by the variations in the median profiles across different mass ranges. This dependency highlights the impact of additional physical processes considered in hydrodynamical simulations on the overall mass distribution.

\section{Methods}\label{sec-3}

\subsection{The deep learning model}
\label{sec-3.1}

In the following, we present the theoretical underpinnings and details of the implementation of the Deep Learning framework. Convolutional Neural Networks (CNNs) have demonstrated significant efficacy in image-based tasks, leveraging multiple hidden layers equipped with convolution and pooling layers for feature extraction \citep{Simonyan2014, krizhevsky2017imagenet}. Notably, CNNs require minimal preprocessing of input data, as the network autonomously learns the convolutional filters \citep{Schmidhuber2014}.

Originally developed for biomedical image segmentation, the U-Net architecture has been successfully adapted for astronomical purposes. For instance, \citet{Akeret2017} utilized a U-Net CNN to filter out radio frequency interference from radio telescope data. Similarly, \citet{Berger2019} employed a three-dimensional U-Net variant, V-Net \citep{Milletari2016}, to forecast and segment dark matter halos in galaxy simulations. Additionally, \citet{Aragon-Calvo2019} employed a V-Net to delineate the cosmological filaments and walls constituting the large-scale structure of the Universe.

\begin{figure*}
\centering
\includegraphics[width=0.9\textwidth]{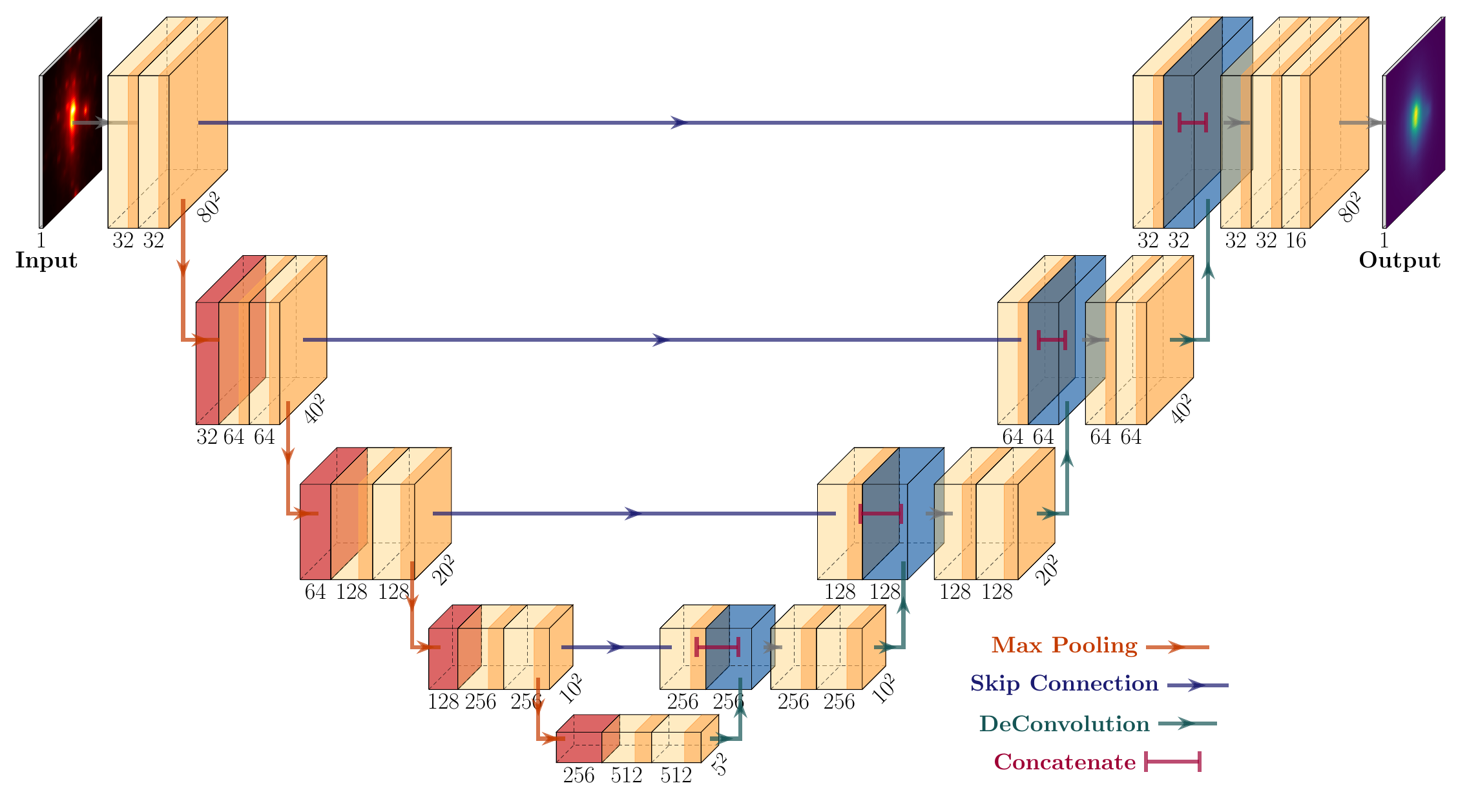}
\caption{ Illustration of the U-Net architecture for observable data prediction. The diagram depicts the network’s components and connectivity. Yellow boxes represent convolutional networks, responsible for extracting features from input volumes. Red boxes indicate network blocks after max-pooling operations, reducing spatial dimensions while preserving feature channels. Blue boxes represent network blocks after deconvolution (or transposed convolution) operations, which upsample the feature maps. The numbers within each box denote the number of filters/neurons, while the number on the right-hand side of each box corresponds to the 2D dimension of the layer. Made with PlotNeuralNet  TeX code \citep{Iqbal2018}.  }
\label{fig:UNET} 
\end{figure*}

The U-Net architecture comprises two principal paths: the contracting path (encoder) and the expansive path (decoder). The encoder captures the image global features by progressively reducing spatial dimensions and increasing feature depth, while the decoder reconstructs image resolution by integrating the global features with localization information. The decoder is in charge of generating the simulated multiview images from the global features. Facilitated by skip connections, the U-Net architecture ensures the capture of both high-level and low-level features.  The U-Net that is being evaluated for our application is shown in Figure \ref{fig:UNET} and is explained below. 

The model incorporates key components such as:

\begin{itemize}
	\item \textbf{Convolutional Neural Networks:} Foundational layers of the U-Net model, leveraging learnable filters to perform convolutions on input data, extracting pertinent features across varying spatial scales \citep[for an introduction, see, e.g.,][]{OShea2015}.
	\item \textbf{Batch Normalization:} Applied post-convolution to normalize activations within each mini-batch, improving the network stability and accelerating the training process \citep{Ioffe2015}.
	\item \textbf{Rectified Linear Unit (ReLU) Activation:} Introduces non-linearity, enabling the network to learn complex patterns and mitigate the vanishing gradient problem \citep{Nair2010}.
	\item \textbf{Max Pooling:} Downsamples reducing the spatial dimensions while retaining the most salient information   \citep{Scherer2010}.
	\item \textbf{Deconvolution (Transpose Convolution):} Upsamples feature maps, recovering the spatial resolution lost during encoding \citep{Zeiler2010}.
	\item \textbf{Copy and Concatenation:} Incorporates skip connections, merging encoder and decoder feature maps to fuse low-level and high-level features.
 \end{itemize}

The following is how these components are added to the architecture:

\begin{enumerate}

\item \textbf{Contracting Path (Encoder):} Consists of multiple blocks of convolutional layers, each followed by batch normalization, ReLU activation, and max pooling for downsampling. This sequential process, with increasing filter sizes (32, 64, 128, 256), reduces spatial dimensions while enhancing feature depth.

\item \textbf{Bottleneck:} Where the feature map undergoes further processing via two convolutional layers (each with 512 filters), batch normalization, ReLU activation, and dropout to mitigate overfitting. The key difference in this path is that the deconvolution starts being applied and is directly concatenated to the last layer of the previous path.

\item \textbf{Expansive Path (Decoder):} Mirrors the contracting path, employing upsampling and concatenation with the encoder feature maps to recover spatial information. This process, with decreasing filter sizes (256, 128, 64, 32, 16), employs transposed convolutional layers until the the size of the output image is reached.

\end{enumerate}

\subsection{Training}
\label{sec-3.2}

\begin{figure*}
\centering
\includegraphics[width=0.85\textwidth]{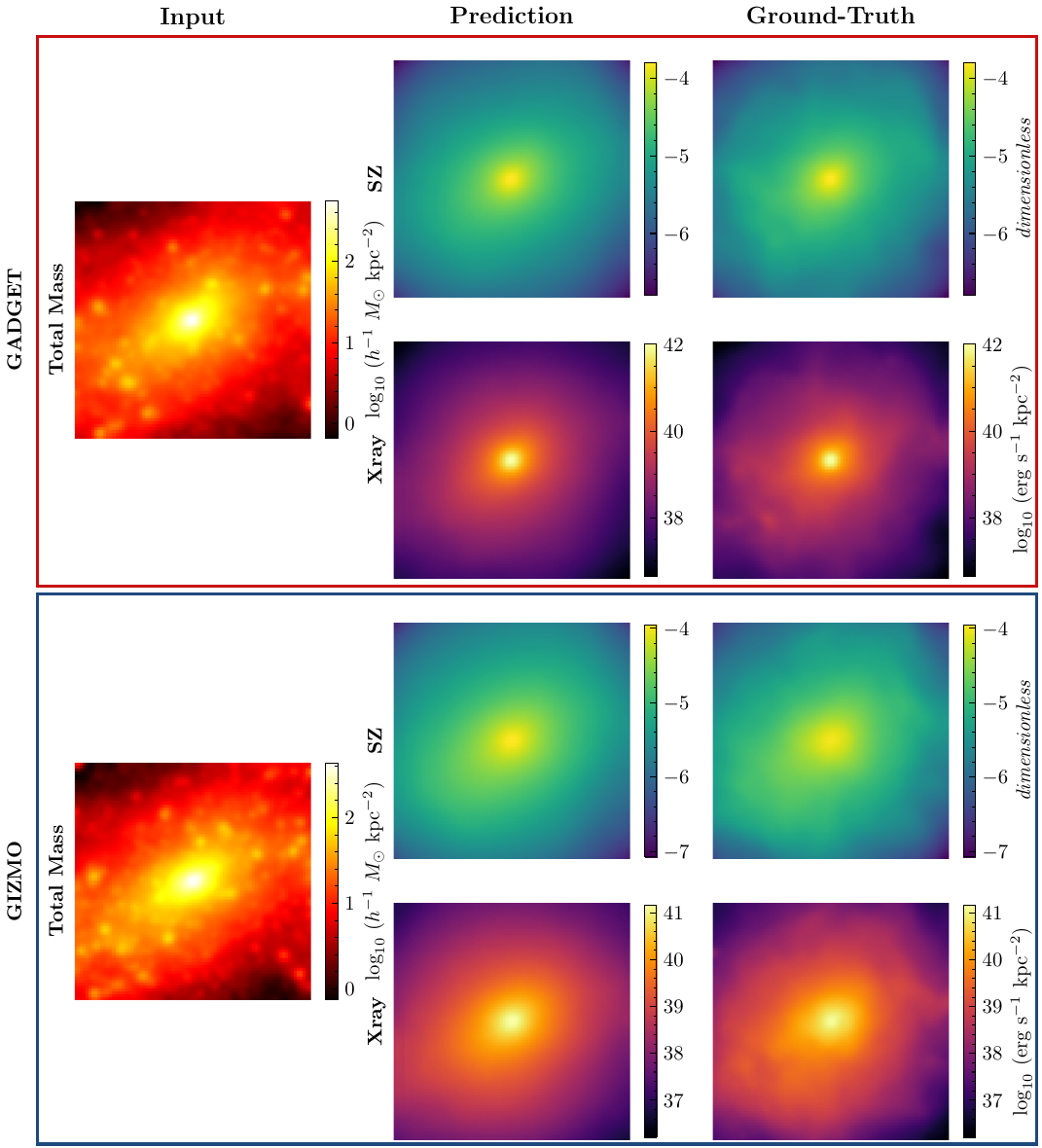}
\caption{Visualization of the input, model predictions, and ground-truth maps for the same halo (HID: 128000000000001) in the same region (45) at redshift 0, using data from two simulations: \gadget{} (red box) and \gizmo{} (blue box). The first column displays the total mass maps used as input. The second column shows the U-Net model predictions in both SZ and X-ray formats. The third column depicts the corresponding ground-truth maps. The model predictions are generated using the U-Net \gizmo+\gadget{} model with a map never seen during training. The maps are shown in log$_{10}$ scale to better highlight the spatial and structural details.}
\label{fig:Predictions_Multiview} 
\end{figure*}

\begin{table}
\begin{tabular}{lll}
\hline
Hyper-parameter    & Description                                                                                         & Value \\ \hline
lr                 & \begin{tabular}[l]{@{}l@{}}learning rate\end{tabular}                                                                                         &  $10^{-3}$     \\
& & \\
epochs             & \begin{tabular}[l]{@{}l@{}}number of epochs\end{tabular}                                                                                        & 100   \\
& & \\
batch\_size        & \begin{tabular}[l]{@{}l@{}}batch size for each step\\ of an epoch\end{tabular}                      & 64    \\
& & \\
optimizer          & \begin{tabular}[l]{@{}l@{}}optimizer for training\end{tabular}                                                                               & Adam  \\
& & \\
regularization     & \begin{tabular}[l]{@{}l@{}}dropout rate\end{tabular}                                                                                        & 0.2   \\
& & \\
loss               & \begin{tabular}[l]{@{}l@{}}loss function that relates \\ true targets with predictions\end{tabular} & MAE   \\
& & \\
reduceLR\_factor   & \begin{tabular}[l]{@{}l@{}}factor for which the \\ learning rate is reduced\end{tabular}            & 0.5   \\
& & \\
reduceLR\_patience & \begin{tabular}[l]{@{}l@{}} number of epochs without \\ improvement until LR \\ adjustment\end{tabular}                     & 5     \\
& & \\
reduceLR\_min      & \begin{tabular}[l]{@{}l@{}}lower bound on the\\  learning rate\end{tabular}                         & $10^{-8}$    \\ \hline
\end{tabular}
\caption{ Hyper-parameters used in the U-Net architecture Figure \ref{fig:UNET}. }
\label{tab:HyperPa}
\end{table}


The training of the U-Net models involved a structured and detailed process, designed to optimize performance and ensure robustness across different datasets. The U-Net's hyperparameters, detailed in Table \ref{tab:HyperPa}, remained fixed throughout the training process, providing a consistent framework for model comparison and evaluation. The model itself encompasses approximately $7.8 \times 10^6$ trainable parameters, allowing for significant flexibility and capacity in learning complex patterns within the data.

Before training, extensive data preprocessing was conducted to refine the input and target datasets for quality and compatibility. This preprocessing includes exploring various normalization techniques to enhance model performance, as elaborated in Appendix \ref{Apx:training_metrics}. Ultimately, the data was normalized using the \textit{Standardization for Log10} technique, which involves applying the logarithmic transformation followed by standard normalization, keeping the dynamic range of the data while preserving its statistical properties. This method involved these two steps:

\begin{enumerate}
	\item Applying the logarithmic transformation $X_{\text{log10}} = \log_{10} (X + \min(X))$.
	\item Standardizing the logarithmically transformed data as $x' = ({x_{\text{log10}} - \mu(X_{\text{log10}})} ) / \sigma(X_{\text{log10}})$.

\end{enumerate}

Here, $X$ represents the dataset, $X_{\text{log10}}$ is the dataset after logarithmic transformation, $x_{\text{log10}}$ is an individual map from $X_{\text{log10}}$. Note that the standardisation involves the mean $\mu$ and standard deviation  $\sigma$ of the whole dataset.

The models were trained on three distinct hydrodynamical datasets: \textsc{Gizmo}+\textsc{Gadget}, \textsc{Gizmo}, and \textsc{Gadget}. Each dataset represents the use of a specific simulation as training data, resulting in two separate networks: SZ and X-ray. Consequently, a total of six individual U-Net models (Table \ref{tab:Models}) were trained to predict observable maps in galaxy clusters. For data splitting, the training dataset comprises 80\% of the samples, while 20\% is allocated to the testing. The training process utilizes the ADAM optimizer \citep{Kingma2014}, with the mean absolute error (MAE) serving as the loss function $L$, measuring the discrepancy between the predicted and ground-truth pixel values. The learning rate is adjusted dynamically, decreasing by a factor of 0.5 if the loss function does not improve after 5 epochs. When the model completes 100 epochs, the training is stopped. The training process for each model took approximately 2-3 hours, leveraging the computational power of an NVIDIA A100-SXM4-40GB GPU. This high-performance GPU facilitated efficient training, enabling the models to process these large datasets rapidly. 

\begin{table}
\centering
\resizebox{\columnwidth}{!}{%
\begin{tabular}{llll}
\hline
Model Notation                       & \gizmo{} dataset  & \gadget{} dataset & Observable \\ \hline
U-Net \textsc{Gizmo}+\textsc{Gadget} & \checkmark & \checkmark & SZ   \& X-ray        \\
U-Net \textsc{Gizmo}                 & \checkmark & -          & SZ   \& X-ray      \\
U-Net \textsc{Gadget}                & -          & \checkmark & SZ   \& X-ray      \\
\end{tabular}%
}
\caption{U-Nets models with their respective training dataset and predicted observable.}
\label{tab:Models}
\end{table}

\section{Results}\label{sec-4}

\begin{figure}
\centering
\includegraphics[width=0.8\columnwidth]{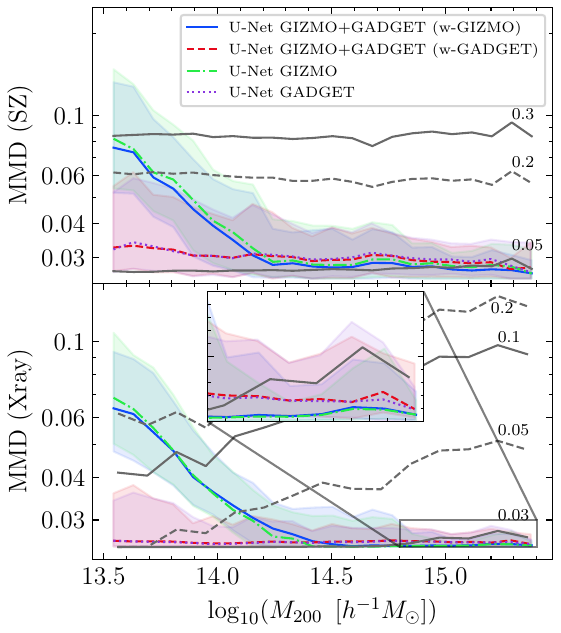}
\caption{ Maximum Mean Discrepancy (MMD), assesses the similarity between two maps, with values closer to 0 indicating greater similarity. SZ maps (top) and X-ray maps (bottom) are subjected to metrics. The noise percentage $\alpha$ that is used to scale the MMD values is represented by the numbers in black on the right side of the grey lines, which indicate the perturbed maps MMDs values as stated by equation (\ref{eq:noise_MMD}). The shaded regions represent the $16^{\text{th}}$ to $84^{\text{th}}$ percentiles, while the lines represent the median metric values. The legend specifies the comparisons between the \gizmo{} and \gadget{} datasets and the outcomes of the U-Net \textsc{Gizmo}+\textsc{Gadget}. These comparisons are assessed independently: (w-\gizmo) and (w-\gadget). }
\label{fig:Test_MMD} 
\end{figure}

In this section, we evaluate the quality of the observable maps that our model predicts. To achieve a preliminary assessment, we compare the ground-truth maps with their corresponding predicted maps in the test set through simple visualizations. This initial analysis is performed using the \gizmo{} and \gadget{} datasets with the U-Net \gizmo+\gadget{} model. Figure \ref{fig:Predictions_Multiview} presents this comparison. In the first column, we show the total mass maps of the same halo for both the \gizmo{} and \gadget{} runs. To the right, we display the predicted observable maps for the SZ and X-ray from the U-Net \gizmo+\gadget{} model. The final column contains the corresponding ground-truth maps. In general, compared to the ground-truth maps, the predicted SZ maps appear smoother, and the predicted X-ray maps exhibit larger numerical differences.

While this visualization in Figure \ref{fig:Predictions_Multiview}  provides an initial insight, it is insufficient for quantitatively assessing the similarity between the predicted and true maps. Thus, we employ a set of additional metrics to measure discrepancies and also compare their scaling relations. To mitigate the risk of numerical errors during metric computations, each observable map is normalized before applying each metric. This normalization is done by dividing each observable map by the mean of the corresponding ground-truth map. The metrics considered are the Maximum Mean Discrepancy (MMD), the Mean Relative Difference (MRD), and Structural Similarity Index Measure (SSIM). These analyses are conducted using the test datasets from the hydrodynamical simulations as well as the dataset from dark matter-only simulations, which are discussed in the following sections.  From now on, we will refer to the ground-truth maps as $X$ and the predicted maps as $Y$.

\subsection{Results on the test dataset} \label{sec-4.1}


 We present the results obtained using the test dataset, which comprises 503 clusters from the \gadget{} simulation and 504 clusters from the \gizmo{} simulation, each subjected to 29 rotations. For the U-Net \gizmo+\gadget{} model, we calculate performance metrics separately for each dataset. When using the \gizmo{} simulation for prediction and metric calculation, we denote this as (w-\gizmo{}). Similarly, when using the \gadget{} simulation, we denote it as (w-\gadget{}). Moreover, to simplify the discussion of scaling relations, we present only the results of the predictions from the U-Net \gizmo+\gadget{} model.

\subsubsection{Metrics}
\label{sec-4.1.1}

\textbf{Maximum Mean Discrepancy (MMD)}, also known as the \textit{two-sample test}, is a statistical measure used to determine whether two distributions, $P_X$ and $P_Y$, are identical based on samples $X$ and $Y$. This metric compares statistics, and a value close to zero indicates that the samples are likely drawn from the same distribution \citep{Gretton2008, Gretton2012}.  Typically employed as a loss function, MMD finds utility in training generative adversarial neural networks \citep{Li2015, KarolinaDziugaite2015}. In our context, we utilize it to assess the fidelity of predicted observable maps. The formal definition of MMD is probabilistic; however, we utilize its unbiased empirical estimate. By selecting a kernel function $k$, two datasets $X$ and $Y$, and the number of samples within them $n$ and $m$ respectively, we compute this metric as follows:

\begin{equation}
    \begin{split}
    \text{MMD}^2(X, Y) = & \frac{1}{n^2} \sum_{i=1}^{n} \sum_{i'=1}^{n} k(x_{i}, x_{i'}) \\
    & - \frac{2}{n m} \sum_{i=1}^{n} \sum_{j=1}^{m}  k(x_{i}, y_{j}) \\
    & \quad + \frac{1}{m^2} \sum_{j=1}^{m} \sum_{j'=1}^{m}  k(y_{j}, y_{j'}),
    \end{split}
    \label{eq:MMD}
\end{equation}

Here, $x_i$ denotes the ground-truth pixels, $y_j$ represents the predicted pixels from our U-Net model, and we have $m = n$ since the maps are $80 \times 80$ pixels. For our application, we employ the Gaussian kernel, defined as:

\begin{equation}
	k(x_i, y_j) = \text{exp} \left( -\frac{ ||x_i - y_j||^2 }{ 2\sigma^2} \right)
\end{equation}

In this equation, $\sigma$ is the bandwidth parameter. To evaluate MMD, a range of $\sigma$ values are typically used. Thus, we consider $\sigma$ values of [0.01, 0.1, 1, 0.25, 0.5, 0.75, 1.0, 2.5, 5.0, 7.5, 10.0, 100.0] for the kernel $k$ and compute the MMD as the maximum value estimated by equation (\ref{eq:MMD}).

Given that MMD lacks an upper bound, interpreting its values in comparison to other metrics can be challenging. To address this, we establish a relative scale for assessing distribution differences by calculating the MMD between the ground-truth mass map $X_i$ and a perturbed map $P_{X_i}$. The perturbed map was generated by adding random Gaussian noise to the ground-truth map. Both maps were normalized by dividing by the mean of the respective ground-truth map. The perturbed map is defined as follows:

\begin{equation} 
P_{X_i} = X_i + \mathcal{N}(0, ({I_{200}}/{N_{200}} \cdot \alpha )^2)
\label{eq:noise_MMD} 
\end{equation}

Here, $I_{200}$ represents the sum of the pixel's values inside $R_{200}$, and $N_{200}$ denotes the number of pixels at $R_{200}$ . The parameter $\alpha$ quantifies the percentage influence on the noise map. To determine representative scales in our MMD results, we tested various values of $\alpha$ as shown in Figure \ref{fig:Test_MMD}, from 0.03 to 0.3.

Next, we computed the MMD between the predicted maps $Y$ and the ground-truth maps $X$ for the three different U-Nets (\gizmo{}+\gadget{}, \gizmo{}, \gadget{}). The median MMD values for SZ and X-ray maps are plotted against the cluster mass for each of the three models for each observable in Figure \ref{fig:Test_MMD}. The grey lines represent the median MMD values of perturbed maps, using  $\alpha = [0.05, 0.2, 0.3]$ for the SZ maps and $\alpha = [0.03, 0.05, 0.1, 0.2]$ for the X-ray maps, providing a relative scale of the MMD.

Analyzing the SZ predictions, the grey lines indicate that U-Net \textsc{Gadget} and U-Net \textsc{Gizmo+\textsc{Gadget}} (w-\textsc{Gadget}) exhibit discrepancies less than 20 \% map noise. However, MMDs fall below 5\% for all models beyond $10^{14.2} h^{{-1}} M_{{\odot}}$. In X-ray predictions, a similar trend emerges but with varying noise percentages. Notably, at 3\% noise, the U-Nets exhibit similar behaviours as the one observed in the SZ predictions, while at 2\% noise, all models demonstrate similar MMDs. In general, these results suggest that lower-mass clusters pose greater prediction challenges for the \textsc{Gizmo} simulation. Additionally, U-Nets predicting X-ray maps outperform those predicting SZ maps.
Furthermore, U-Net \textsc{Gizmo+\textsc{Gadget}} differentiates between input mass maps from \textsc{Gizmo} and \textsc{Gadget} simulations.

\begin{figure}
\centering
\includegraphics[width=0.8\columnwidth]{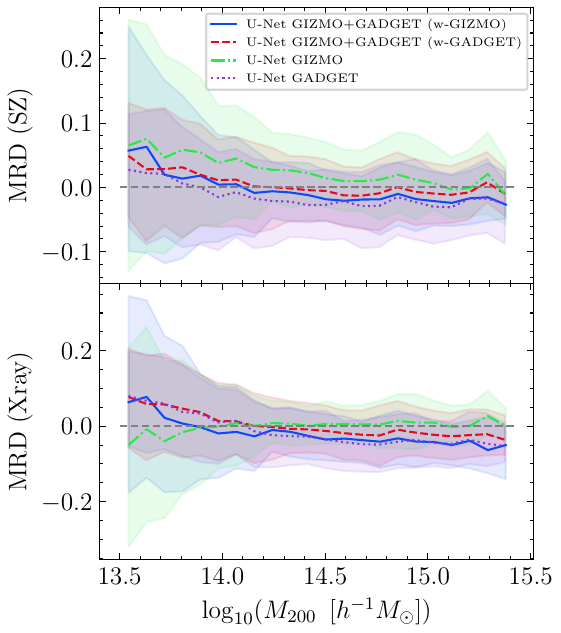}
\caption{ The difference between the predicted and ground-truth maps measured by the Mean Relative Difference (MRD), as shown in equation (\ref{eq:MRD}). The grey dashed line at 0 indicates identical predicted and ground-truth maps. This metric is applied to SZ maps (top) and X-ray maps (bottom). The median metric values are represented by the lines, and the shaded portions correspond to the $16^{\text{th}}$ to $84^{\text{th}}$ percentiles. We show the U-Net \textsc{Gizmo}+\textsc{Gadget} evaluated separately (w-\gizmo{}, in blue) and (w-\gadget{}, in red ) as well as the performance of individual models U-Net \textsc{Gizmo} in green and U-Net \textsc{Gadget} in purple colour.}
\label{fig:Test_MRD} 
\end{figure}

\textbf{Mean Relative Difference (MRD)}, also known as \textit{arithmetic mean change}, is a quantitative measure that indicates the disparity between two values, providing a ratio that remains unaffected by the units employed \citep[see][Chapter 1]{Vartia1976}. In our context, we utilize this metric to compare the pixels between the ground-truth maps and the model predictions, specifically to avoid computational issues with zero values. The MRD is calculated using the following equation:

\begin{equation}
    \text{MRD}(X, Y) = \frac{1}{N^2} \sum_{i=1}^{N} \sum_{j=1}^{N} \frac{2 (Y_{ij} - X_{ij})}{Y_{ij} + X_{ij}},
    \label{eq:MRD}
\end{equation}

where $N$ represents the dimension of the maps in pixels, this metric is particularly useful for determining whether the data is biased, indicating if the predictions are overestimated or underestimated. In our case, the MRD has specific boundaries due to the non-negative nature of the values in the observable maps. Values approaching 2 or -2 indicate a substantial discrepancy between the pixels, whereas a value of 0 denotes no difference between the compared values. However, interpreting MRD values is not as straightforward as interpreting metrics such as percentage difference ($100 \times (Y-X)/X$). We have estimated that MRD values between -0.3 and 0.3, a scale similar to the percentage difference. This suggests that the MRD and percentage difference metrics are comparable for discrepancies below 30\%.

In Figure \ref{fig:Test_MRD}, we present the MRD for the SZ and X-ray maps as a function of cluster mass $M_{200}$. The median MRD values are depicted as lines, while the shaded regions represent the $16^{\text{th}}$ to $84^{\text{th}}$ percentiles. The analysis reveals similar trends in predictions for both observables, with an exception noted in the U-Net \gizmo{} predictions. Specifically, the median of the U-Net \gizmo{} predictions indicates that SZ maps are predominantly overestimated across almost all masses, whereas the X-ray maps remain mostly stable around zero. For other predictions, the median values indicate overestimation of observable maps for masses below $10^{14} h^{-1} M_{\odot}$, followed by underestimation beyond this mass threshold. Furthermore, the shaded regions reveal that the predictions typically do not exceed a 20\% difference from the ground-truth maps, with the \gadget{} predictions demonstrating greater accuracy (scatter less than 10\%) compared to the \gizmo{} predictions. This results aligns with the previous results obtained from the MMD analysis. It is also important to note that MRD slightly oscilates around zero, meaning a good agreement between predicted and ground-truth maps.

\begin{figure}
\centering
\includegraphics[width=0.8\columnwidth]{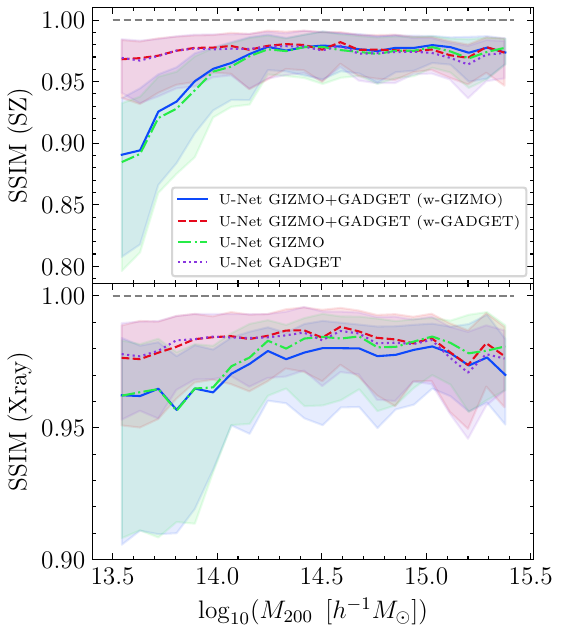}
\caption{ The Structure Similarity Index Measure (SSIM) for SZ maps (top) and X-ray maps (bottom) calculates the similarity between the ground-truth and predicted maps using equation (\ref{eq:MRD}). The highest structural similarity is indicated by the grey dashed line at 1. The lines show the median metric values, while the shaded areas indicate the $16^{\text{th}}$ to $84^{\text{th}}$ percentiles. The comparison of The U-Net \textsc{Gizmo}+\textsc{Gadget} was evaluated separately (w-\gizmo{}) and (w-\gadget{}) and is presented in the legend together with the comparisons with the \gizmo{} and \gadget{} U-Nets. }
\label{fig:Test_SSIM} 
\end{figure}

\textbf{Structure Similarity Index Measure (SSIM)} is a method used to evaluate the similarity between two images by leveraging the premise that pixels have strong dependencies that convey significant information about the structure, particularly when they are spatially proximate \citep{Wang2004}. This metric focuses on assessing and comparing the luminance, contrast, and structure of the two images using a specified pixel window that slides through the images. In this study, we utilize the Python package Scikit-Image\footnote{\url{https://scikit-image.org/docs/stable/}} \citep{vanderWalt2014}, which employs a default window size of $7 \times 7$ pixels. The SSIM is calculated using the simplified\footnote{The definition of the SSIM index from \cite{Wang2004} is a more complex expression with multiple parameters.} expression:

\begin{equation}
    \text{{SSIM}}(X, Y) = \frac{1}{M^2} \sum_{i=1}^{M} \frac{{(2\mu_{x_i}\,\mu_{y_i} + c_1)(2\,\sigma_{{x_i}{y_i}} + c_2)}}{{(\mu_{x_i}^2 + \mu_{y_i}^2 + c_1)(\sigma_{x_i}^2 + \sigma_{y_i}^2 + c_2)}}
    \label{eq:SSIM}
\end{equation}

Here, $M$ represents the number of local windows in the image, $\mu_{x_i}$ and $\mu_{y_i}$ represent the mean intensities of the sliding window on the respective maps, $\sigma_{x_i}$ and $\sigma_{y_i}$ denote the standard deviations of these windows, and $\sigma_{x_i y_i}$ is the covariance between corresponding windows in the ground-truth maps and the model predictions. The constants $c_1 = 0.01 \times (\max(X) - \min(X))$ and $c_2 = 0.03 \times (\max(X) - \min(X))$\footnote{The SSIM function in the Scikit-Image package uses these default values of 0.01 and 0.03, which were derived from \cite{Muller2020}}  are included to prevent instability in the calculations. 

The SSIM for the SZ and X-ray maps as a function of the cluster mass $M_{200}$ is shown in Figure \ref{fig:Test_SSIM}. The median values are shown as lines, and the shaded regions represent the $16^{\text{th}}$ to $84^{\text{th}}$ percentiles. The SSIM index is a number between -1 and 1, where -1 denotes complete anti-correlation, 1 represents perfect similarity and 0 indicates no similarity. Upon analyzing the figure, we find that for the \gizmo{} runs, clusters with masses less than $10^{14.2} h^{-1} M_{\odot}$ had lower SSIM values for the SZ maps. Beyond this mass threshold, the predictions start to match the trend shown in the \gadget{} runs more closely. A similar pattern is noted in the SSIM values for the X-ray maps, although the effect is less pronounced. Overall, the SSIM values suggest that the predicted X-ray maps exhibit slightly better similarity compared to the SZ maps. This observation is consistent with the general trend that X-ray map predictions are more accurate than SZ map predictions. 

\subsubsection{Scaling Relations}

\begin{figure}
\centering
\includegraphics[width=0.9\columnwidth]{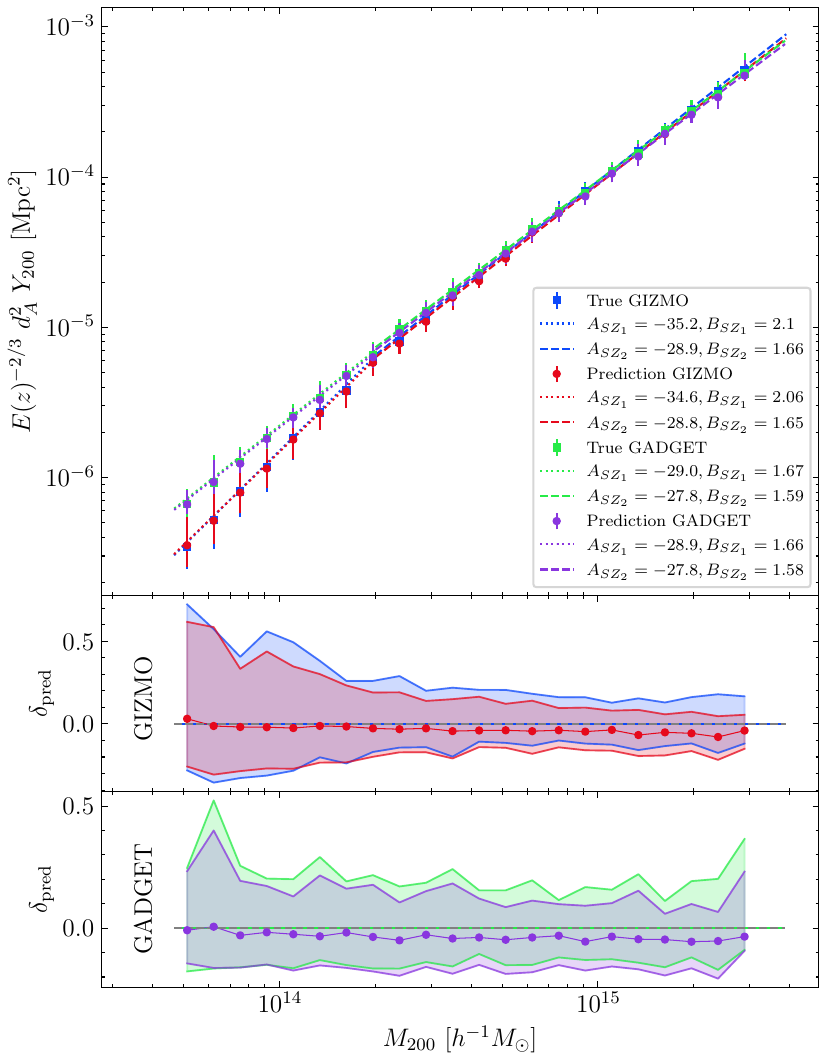}
\caption{Scaling relation between the integrated Compton parameter $Y$ and the mass $M$ for the U-Net \textsc{Gizmo}+\textsc{Gadget} (Table \ref{tab:Models}) within the radius $R_{200}$. The squares represent the median value for the true targets (\textsc{Gizmo}: blue, \textsc{Gadget}: green), while the circles indicate the median value for the predictions (\textsc{Gizmo}: red, \textsc{Gadget}: purple). The error bars correspond to the $16^{\text{th}}$ to $84^{\text{th}}$ percentile range. On  the top graph, the lines depict the fitted results obtained from equation (\ref{eq:SZ-fit}).  Given that the pivot point in steepness occurs at approximately $2\times 10^{14} h^{-1} M_\odot$, the dotted lines and the $SZ_1$ subscripts represent the fitted results for masses below this threshold, while the dashed lines and the $SZ_2$ subscripts represent the fitted results for masses above it. The bottom panels shows the median (solid lines), the $16^{\text{th}}$  and $84^{\text{th}}$ percentiles (shaded region) of the relative error $\delta_{\text{pred}} = (Y_{\text{pred}} - Y_{\text{true}}) / Y_{\text{true}}$. }
\label{fig:SZ_scaling_train}
\end{figure}

\begin{figure}
\centering
\includegraphics[width=0.9\columnwidth]{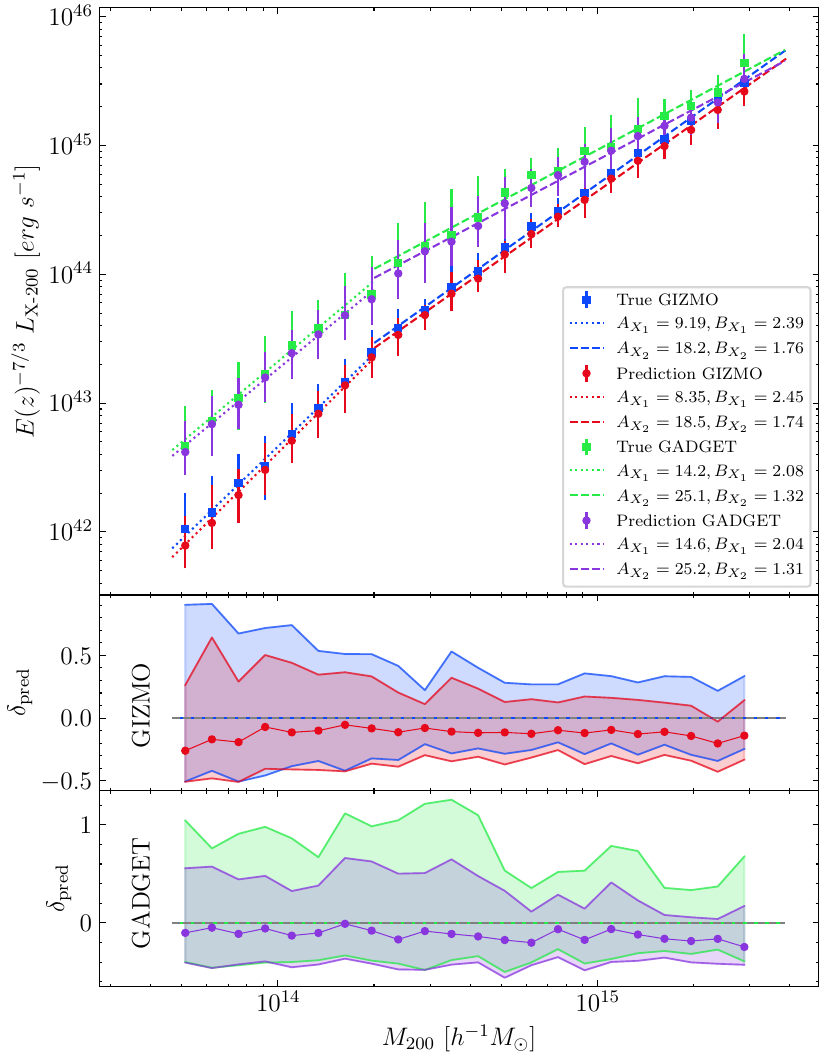}
\caption{ Relationship between the X-ray bolometric luminosity $L_X$ and the mass $M$ for the U-Net \textsc{Gizmo}+\textsc{Gadget} (Table \ref{tab:Models}) within the radius $R_{200}$. The squares represent the median value for the true targets (\textsc{Gizmo}: blue, \textsc{Gadget}: green), while the circles indicate the median value for the predictions (\textsc{Gizmo}: red, \textsc{Gadget}: purple). The error bars correspond to the $16^{\text{th}}$ to $84^{\text{th}}$ percentile range. On the top graph, the lines are fitted according to equation (\ref{eq:Xray-fit}). Given that the pivot point in steepness occurs at approximately $2\times 10^{14} h^{-1} M_\odot$, the dotted lines and the $X_1$ subscripts represent the fitted results for masses below this threshold, while the dashed lines and the $X_2$ subscripts represent the fitted results for masses above it. The bottom panels shows the median (solid lines), the $16^{\text{th}}$  and $84^{\text{th}}$ percentiles (shaded regions) of the relative error $\delta_{\text{pred}} = (L_{X\text{-pred}} - L_{X\text{-true}}) / L_{X\text{-true}}$ . }
\label{fig:Xray_scaling_train}
\end{figure}

\begin{figure*}
\includegraphics[width=0.8\textwidth]{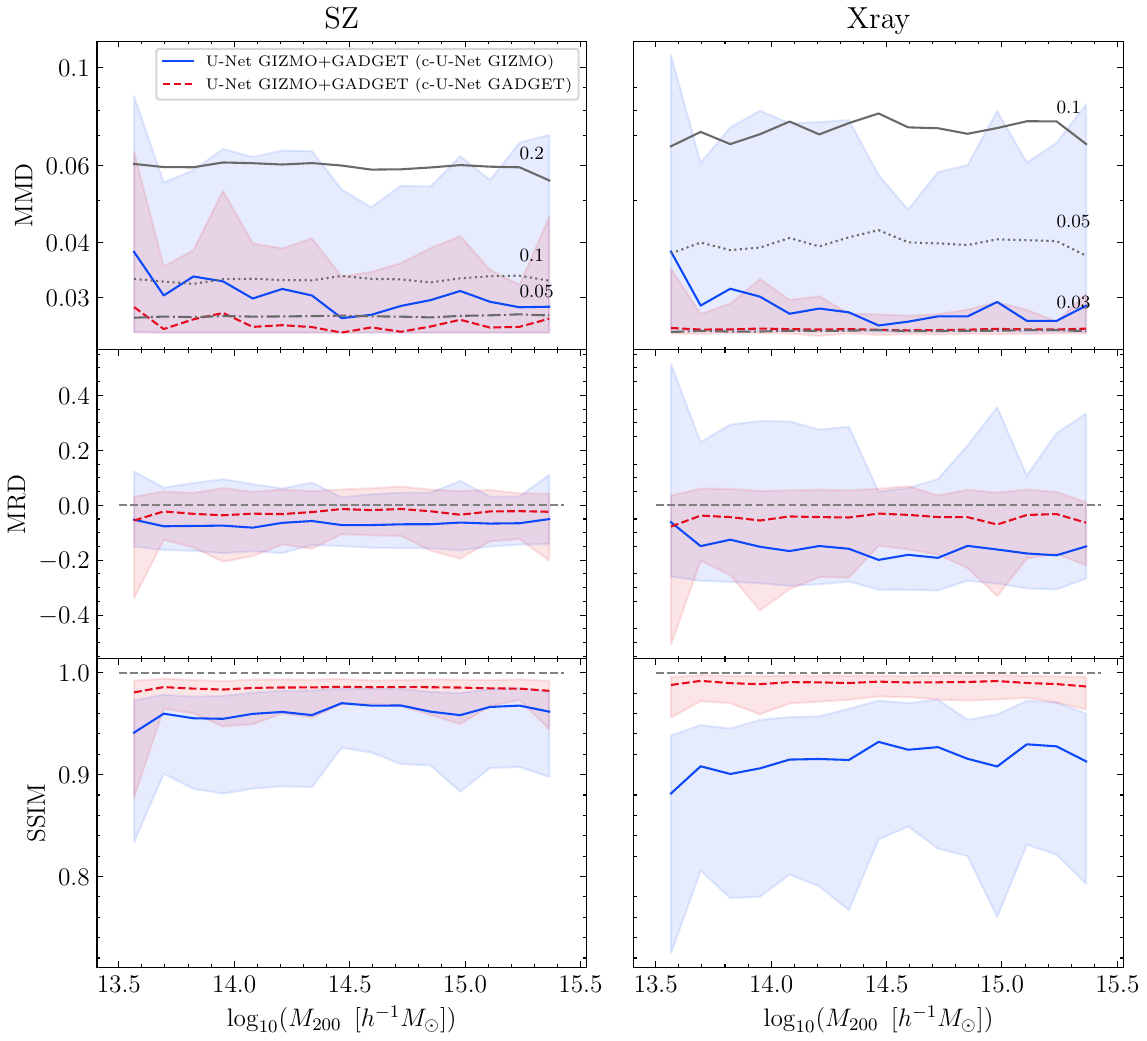}
\caption{ Results of the U-Net trained with \textsc{Gizmo}+\textsc{Gadget} compared with the \textsc{Gizmo} and \textsc{Gadget} datasets separately  when generating observational maps from {\sc The300} \textsc{Gadget}-X DM-only simulations. We denote in the legend their respective comparisons: (c-U-Net \gizmo{}) and (c-U-Net \gadget{}) The first metric, Maximum Mean Discrepancy (MMD), assesses similarity between maps, with values closer to 0 denoting greater resemblance. The second metric measures mean relative difference, where values nearing 0 indicate heightened similarity. Lastly, the Structural Similarity Index Measure (SSIM) evaluates morphological likeness, with values closer to 1 signifying greater similarity. Metrics are applied to SZ maps (left column) and X-ray maps (right column). Lines represent median values, while shaded regions denote the $16^{\text{th}}$ to $84^{\text{th}}$ percentiles.  }
\label{fig:MetricsComparison} 
\end{figure*}


A conventional approach for determining cluster masses involves establishing relation between total mass and an observable parameter, which is typically calibrated with either simulations or other more convenient observations. Such measurements are typically obtained from X-ray luminosity, gravitational lensing, and the Sunyaev-Zel'dovich effect. A common approach to address this is the use of scaling relations \citep[e.g.,][]{Planck2016SZcosmo,Lovisari2022}. These mass–observable scaling relations are crucial for inferring the masses of galaxy clusters, which play a very important role in cosmology.

Therefore, an important source of uncertainty is the slope and normalization of mass-observable relations \citep[e.g.,][]{Rozo2010, Bocquet2015}. Despite recent developments, several aspects remain not fully understood. These aspects include the origin of intrinsic scatter in scaling relations, their applicability to low-mass clusters and groups, and their variability to redshifts \citep{Henden2019}. Given that hydrodynamical cosmological simulations can aid in understanding these issues, we focus on examining specific relationships for each component. By comparing the scaling relations of predicted maps against known ground-truth maps, we can quantitatively measure the fidelity of the predictions. 

For the SZ maps, we evaluate the scaling relation between the integrated Compton-$y$ parameter $Y_{200}$ and the mass $M_{200}$. This is achieved by applying the method of least squares to fit the following power-law relationship:

\begin{equation} 
d_A^2 Y_{200} = 10^{A_{SZ}} E(z)^{2/3} (M_{200})^{B_{SZ}}.
\label{eq:SZ-fit} 
\end{equation}

Here, $d_A$ denotes the angular diameter distance, $E(z) = H(z)/H_0$ represents the dimensionless Hubble parameter, and $A_{SZ}$ and $B_{SZ}$ are the parameters to be fitted. Figure \ref{fig:SZ_scaling_train} illustrates this relationship by comparing the predictions from U-Net \gizmo{}+\gadget{} with the actual data from the \gizmo{} and \gadget{} simulations. The figure presents the median cylindrical integrated values of $Y$ up to $R_{200}$. Given that the scaling for both observables has a pivot point in steepness at approximately $2\times 10^{14} h^{-1} M_\odot$, we plot the dotted lines representing the fitted results for masses below this threshold, while the dashed lines represent the fitted results for masses above it. The error bars represent the $16^{\text{th}}$ to $84^{\text{th}}$ percentile range. In addition, we display the fitted parameters of equation (\ref{eq:SZ-fit}) as dashed lines, with the corresponding fit values detailed in the legend. Upon examining the figure, it becomes evident that the U-Net \gizmo{}+\gadget{} can separately predict the outcomes of both simulations on which it was trained. This is particularly noticeable at low masses, where the simulations exhibit more significant differences. We calculate the percentage errors for all the scaling parameters as $100 \times |A_{\text{pred}}-A_{\text{true}}|/|A_{\text{true}}|$ (see tables in Appendix \ref{Apx:parameter_tables}). Prior to the pivot point, the errors do not exceed (2.1 ± 0.6)\%, while subsequent to the pivot point, the errors do not exceed (0.5 ± 0.1)\%.

Similarly, for the X-ray maps, we examine the relationship between the X-ray bolometric luminosity $L_{X\text{-}200}$ and the mass $M_{200}$ by fitting the following power-law relationship:

\begin{equation} 
L_{X\text{-}200} = 10^{A_{X}} E(z)^{7/3} (M_{200})^{B_{X}}. 
\label{eq:Xray-fit} 
\end{equation}

In this equation, $A_{X}$ and $B_{X}$ are the parameters to be estimated. This relationship is illustrated in Figure \ref{fig:Xray_scaling_train}, where we compare the U-Net \gizmo{}+\gadget{} predictions with the actual data from the \gizmo{} and \gadget{} simulations. Similar to the previous analysis, the figure presents the median cylindrical integrated values of $L_X$ up to $R_{200}$. Given that the scaling for X-ray luminosity has a pivot point in steepness at approximately $2\times 10^{14} h^{-1} M_\odot$, we plot the dotted lines representing the fitted results for masses below this threshold, while the dashed lines represent the fitted results for masses above it. The error bars represent the $16^{\text{th}}$ to $84^{\text{th}}$ percentile range. Additionally, we display the parameters of the fit of equation (\ref{eq:Xray-fit}) in the legend. Notably, the U-Net \gizmo{}+\gadget{} can also separately predict the outcomes of both simulations on which it was trained. This separation is even more noticeable than in the SZ scaling relation due to the different parameters of both simulations. The percentage errors of the scaling parameters do not exceed (9.1 ± 2.7)\% before the pivot point and the errors do not exceed (1.6 ± 0.3)\% after the pivot point.

\subsection{Applying to DM-only simulation}\label{sec-4.2}

In addition to testing our model, we further assess its performance using the \gadgetx{} DM-only from the \thethree{} project. This serves as the primary test for evaluating the applicability of our model. As noted in Section \ref{sec-2.3}, we selected 2155 halos and chose three perpendicular rotations for each halo. Our presentation of the results follows a similar approach to that outlined in Section \ref{sec-4.1}. We compute the metrics by comparing the predictions of the U-Net \gizmo+\gadget{} model to those of the individual simulation U-Nets. Specifically, we use the notation (c-U-Net \gizmo{}) for comparisons with the predicted observables of the U-Net \gizmo{} and (c-U-Net \gadget{}) for comparisons with the predicted observables of the U-Net \gadget{}. Moreover, to fully capture the discussion on scaling relations, we present the outcomes of the predictions made by all six U-Net models. The scaling relations when using the \music{} DM-only dataset (Section \ref{sec-2.3}) are presented in Appendix \ref{Apx:DMonly_MUSIC}.

\subsubsection{Metrics}
\label{sec-4.2.1}

We employ the same set of metrics used in the test results: maximum mean discrepancy (MMD), mean relative difference (MRD), and structural similarity index measure (SSIM). The results are displayed in Figure \ref{fig:MetricsComparison}, with two columns: the left for the SZ maps and the right for the X-ray maps, each row representing one of the metrics.

The figure illustrates, in general, that the U-Net \gizmo{}+\gadget{} predictions align more closely to the U-Net \gadget{} predictions than the U-Net \gizmo{} predictions. Specifically, the MMD relative scale uses the median of the U-Net \gizmo{} perturbed maps and the U-Net \gadget{} perturbed maps (Eq. \ref{eq:noise_MMD}) with parameters $\alpha = [0.05, 0.1, 0.2]$ for the SZ maps and $\alpha = [0.03, 0.05, 0.1]$ for the X-ray maps. The MMD shows that the X-ray maps (with a scatter of less than 10\% of the noise maps) have fewer probabilistic discrepancies compared to the SZ predictions (which have a scatter close to 20\% of the noise maps). The MRD metric results indicate that in both cases of U-Net \gizmo{} and U-Net \gadget{}, the U-Net \gizmo{}+\gadget{} generally underestimates the observable maps, with a difference of no more than ~20\% for the SZ maps and no more than ~30\% for the X-ray maps. Regarding the SSIM metric, the U-Net \gizmo{}+\gadget{} X-ray predictions show slightly greater similarity to the U-Net \gadget{} ($\text{SSIM}_{\text{median}} = 0.990^{+0.007}_{-0.020}$) than the SZ predictions ($\text{SSIM}_{\text{median}} = 0.985^{+0.008}_{-0.026}$).

Altogether, the metrics results indicate that the predictions of U-Net \gizmo{} +\gadget{} using the total mass data of \gadget{}-X DM-only are very similar to those from the U-Net \gadget{} using the same total mass data. This suggests that the model trained on both simulations effectively recognizes the differences between the total mass distributions. This finding is consistent with the results presented in Section \ref{sec-4.1}, where the same U-Net model successfully differentiates between the simulations using the test dataset.

\subsubsection{Scaling Relations}

\begin{figure}
\centering
\includegraphics[width=0.9\columnwidth]{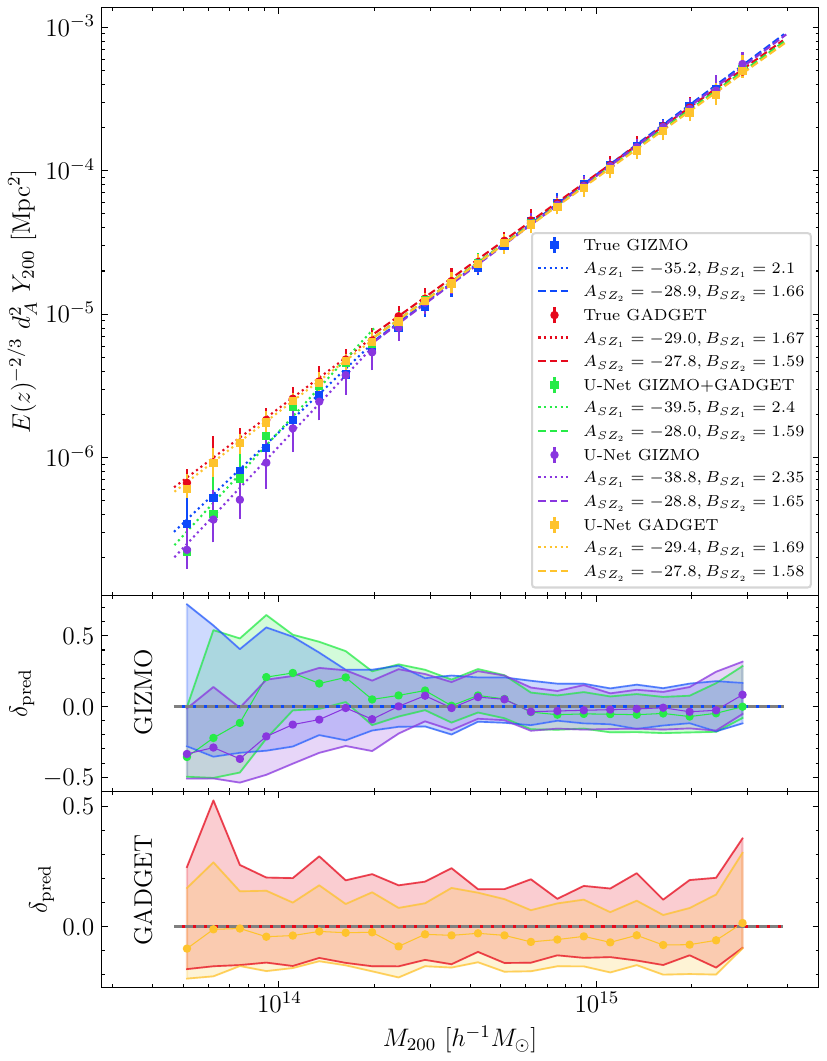}
\caption{ The $Y$–$M$ scaling relation derived from the \textsc{Gizmo} (blue) and \textsc{Gadget} (red) simulations, alongside the relation from the simulated observations generated using U-Nets: \textsc{Gizmo}+\textsc{Gadget} (green), \textsc{Gizmo} (purple), and \textsc{Gadget} (yellow). These scaling relations were obtained using total mass maps from the \textsc{Gadget-X} DM-only dataset as input. The error bars correspond to the $16^{\text{th}}$ to $84^{\text{th}}$ percentile range. On the top graph, the lines depict the fitted results obtained from equation (\ref{eq:SZ-fit}).  Given that the pivot point in steepness occurs at approximately $2\times 10^{14} h^{-1} M_\odot$, the dotted lines and the $SZ_1$ subscripts represent the fitted results for masses below this threshold, while the dashed lines and the $SZ_2$ subscripts represent the fitted results for masses above it. The bottom panels shows the median (solid lines), the $16^{\text{th}}$  and $84^{\text{th}}$ percentiles (shaded region) of the relative error $\delta_{\text{pred}} = (Y_{\text{pred}} - Y_{\text{true}}) / Y_{\text{true}}$. For the U-Net \textsc{Gizmo}+\textsc{Gadget} relative error, we consider the \textsc{Gizmo} simulation as the $Y_{\text{true}}$.}
\label{fig:SZ_FullDMonly_M200} 
\end{figure}

\begin{figure}
\centering
\includegraphics[width=0.9\columnwidth]{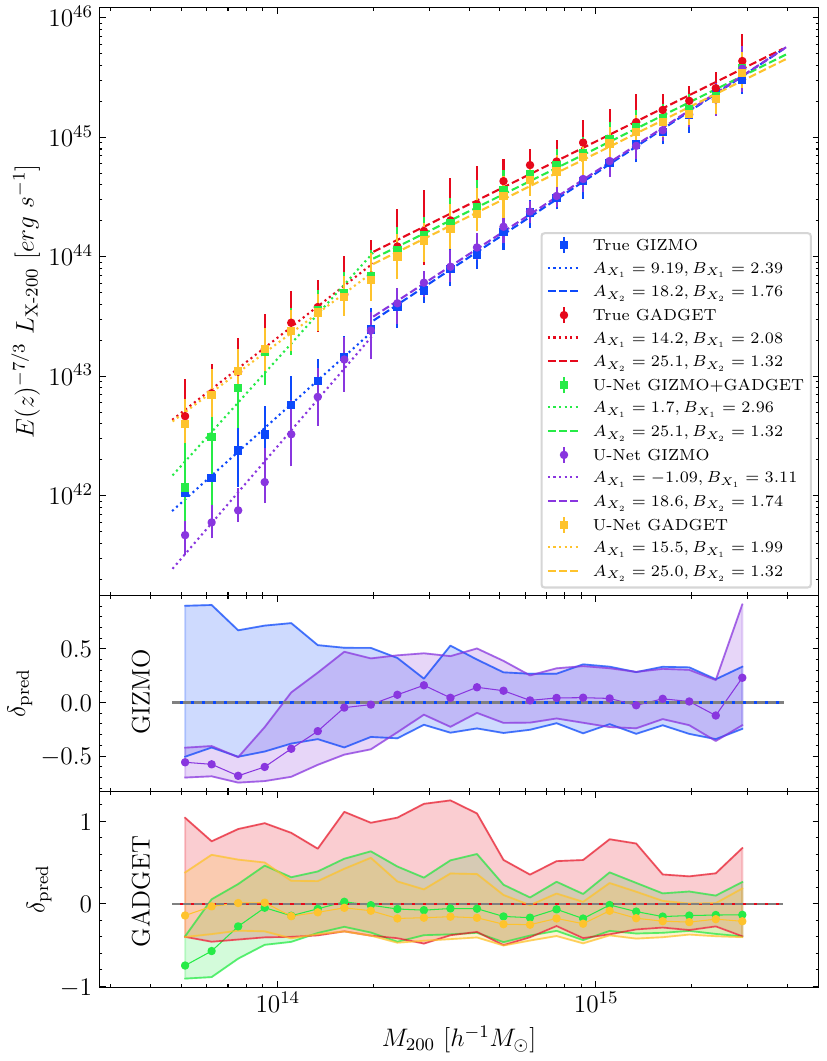}
\caption{ The $L_X$–$M$ relationship derived from the \textsc{Gizmo} (blue) and \textsc{Gadget} (red) simulations, alongside the relation from the simulated observations generated using U-Nets: \textsc{Gizmo}+\textsc{Gadget} (green), \textsc{Gizmo} (purple), and \textsc{Gadget} (yellow). These scaling relations were obtained using total mass maps from the \textsc{Gadget-X} DM-only dataset as input. The error bars correspond to the $16^{\text{th}}$ to $84^{\text{th}}$ percentile range. On the top graph, the lines are fitted according to equation (\ref{eq:Xray-fit}). Given that the pivot point in steepness occurs at approximately $2\times 10^{14} h^{-1} M_\odot$, the dotted lines and the $X_1$ subscripts represent the fitted results for masses below this threshold, while the dashed lines and the $X_2$ subscripts represent the fitted results for masses above it. The bottom panels shows the median (solid lines), the $16^{\text{th}}$  and $84^{\text{th}}$ percentiles (shaded regions) of the relative error $\delta_{\text{pred}} = (L_{X\text{-pred}} - L_{X\text{-true}}) / L_{X\text{-true}}$. For the U-Net \textsc{Gizmo}+\textsc{Gadget} relative error, we consider the \textsc{Gadget} simulation as the $L_{X\text{-true}}$.}
\label{fig:Xray_FullDMonly_M200} 
\end{figure}

We evaluate the scaling relations using equations (\ref{eq:SZ-fit}) and (\ref{eq:Xray-fit}). Figures \ref{fig:SZ_FullDMonly_M200} and \ref{fig:Xray_FullDMonly_M200} depict the signal-mass relation for the SZ and X-ray maps predicted by the three U-Nets: \gizmo{}+\gadget{}, \gizmo{}, and \gadget{}, and the percentage errors for the scaling-law parameters are presented in Appendix \ref{Apx:parameter_tables}.

In Figure \ref{fig:SZ_FullDMonly_M200}, the median values of the cylindrical integrated values of $Y$ up to $R_{200}$ are shown, with error bars representing the percentiles from the $16^{\text{th}}$ to $84^{\text{th}}$. The dashed lines represent the least squares fitted scaling relations, with parameters indicated in the legend. Analyzing the figure, we observe that at low masses, the predicted scalings differ more significantly between the models. When comparing the fitted parameters of the U-Net \gizmo{} and the U-Net \gadget{} to their respective ground-truth datasets, it is observed that the U-Net \gadget{} predicts the $Y$–$M$ scaling relation more accurately, with percentage errors not exceeding (1.5 ± 0.7)\% before the pivot point and (0.2 ± 0.2)\% after the pivot point. For the combined U-Net \gizmo{}+\gadget{}, we determine the percentage error by comparing it with the ground-truth datasets exhibiting the most similar trends. Prior to the pivot point, the trend aligns closely with \gizmo{}, with percentage errors not exceeding (14.4 ± 1.1)\%, while post-pivot point, the trend aligns more closely with \gadget{}, with percentage errors not exceeding (0.5 ± 0.1)\%.

Figure \ref{fig:Xray_FullDMonly_M200} shows the scaling relations for the X-ray maps predicted by the three U-Nets. The median values of $L_{X,200}$ are represented with error bars indicating the percentiles from $16^{\text{th}}$ to $84^{\text{th}}$. The dashed lines are the least-squares fitted scaling relations, with parameters detailed in the legend. From the figure, we note that the predictions of the U-Net \gizmo{} are underestimated compared to the \gizmo{} ground-truth before the pivot point of $2 \times 10^{14}h^{-1}  M_{\odot}$, after which they exhibit a better agreement. By contrast, the predictions of the U-Net \gadget{} align well with the scaling relation of the \gadget{} ground-truth data in the whole mass range. Furthermore, the predictions of the U-Net \gizmo{}+\gadget{} tend closer to the scaling of the \gadget{} run after the pivot point. In contrast, before this pivot point, at very low mass clusters they tend closer to the \gizmo{} run before gradually transitioning towards the \gadget{} run. Both figures suggest that the predictions from using the \gadget{}-X DM-only dataset are more precise for the U-Net \gadget{} model compared to the U-Net \gizmo{} model, particularly for low-mass clusters. This difference may be attributed to the baryonic interactions in the \gizmo{} simulation, which produce clusters where the total mass significantly differs from the ones in the \gadget{}-X DM-only simulations.

Overall, the scaling relations indicate that, although the U-Nets' predictions generally exhibit a scaling similar to the ground-truth datasets, the outcomes are highly dependent on the training data as well as the input data used for prediction. Although the predictions are robust, it is important to note that the models were trained using total matter maps from hydrodynamical simulations. Thus, when predicting using the total mass of DM-only simulations, the relationships involving baryonic effects present in hydrodynamical simulations are not accounted for. However, given that gravitational effects dominate at large scales, the DM-only predictions remain highly accurate for massive clusters where the self-similar model especially holds.

\subsection{Interpretations}
\label{sec-4.3}

\begin{figure*}
\centering
\includegraphics[width=0.9\textwidth]{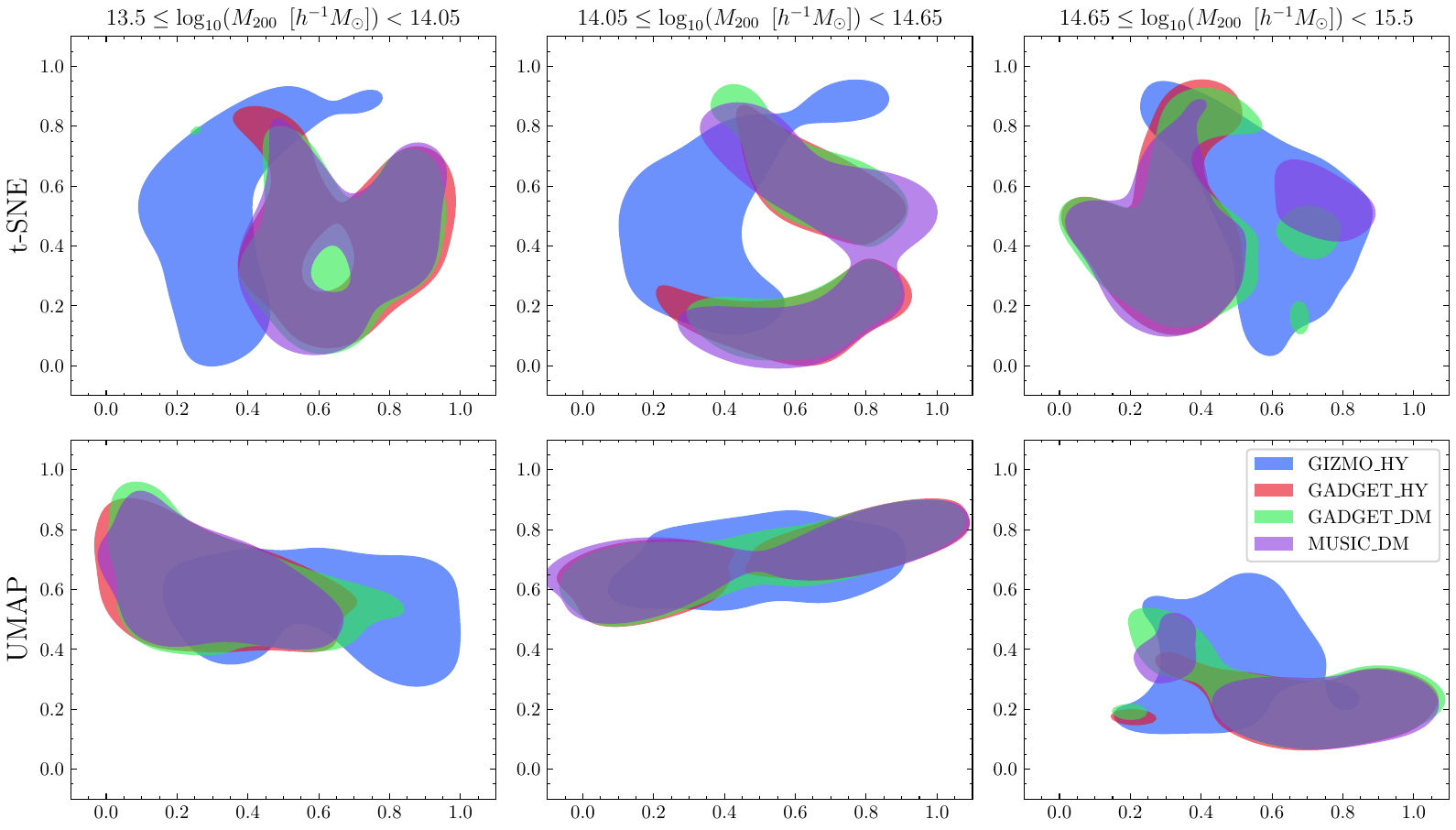}
\caption{ Bi-dimensional visualizations that showcase the final convolutional layer of the SZ U-NET \gizmo+\gadget{} model. These visualizations stem from four distinct input simulations: two hydrodynamical simulations, denoted as \textsc{Gizmo\_hy} and \textsc{Gadget\_hy}, and two N-body dark matter-only simulations, labelled as \textsc{Gadget\_dm} and \textsc{Music\_dm}. The top row illustrates the dimensionality reduction achieved through the application of the probabilistic technique of t-distributed Stochastic Neighbor Embedding (t-SNE), while the bottom row showcases the dimensionality reduction using the Uniform Manifold Approximation and Projection for Dimension Reduction (UMAP) method. An explanation of the contour generation process is detailed in Section \ref{sec-4.3}. Each graph, from left to right, corresponds to the three intervals specified in the plot titles.  }
 \label{fig:tSNE_UMAP}
\end{figure*}

Understanding the behaviour of the U-Net model requires a thorough examination of the inherent differences between the considered datasets. Figure \ref{fig:4sim_density_profiles} shows the normalised total mass profiles of the simulations employed throughout this study, noting disparities with a maximum difference of ~8\% between \gizmo{}\_\textsc{hy} and \gadget{}\_\textsc{hy} mainly in the core and independent of the mass. This variation potentially contributes to the differences observed in the metrics of the predicted maps discussed in Section \ref{sec-4.2.1} and as noted in Appendix \ref{Apx:Gas_profiles} the gas distribution may be the biggest factor, thus indicating that the baryonic processes in \gizmo{} simulation impact directly in the total mass distribution of the clusters. The AGN feedback is more efficient in the \gizmo{} simulation where a noticeable effect is that the gas is blown outside the virial radius \citep{Sorini2022:AGNgizmo, Yang2024:AGNgizmo}.

To further understand the U-Net model, we applied a couple of dimensionality reduction algorithms and visualized the results in a two-dimensional plane. Specifically, we used the following algorithms:

\textbf{t-Distributed Stochastic Neighbor Embedding (t-SNE)} is an algorithm developed by \cite{Maaten2008}. It constructs a probability distribution over pairs of high-dimensional data points, where similar data points have a high probability of being chosen and dissimilar data points have a low probability. This approach is a variation of Stochastic Neighbor Embedding \citep[SNE,][]{Hinton2002}, which considers a conditional probability $p_{j|i}$ given by:


\begin{equation}
   p_{j|i} = \frac{\exp(-|| x_i - x_j ||^2 / 2\sigma_i^2)}{\sum_{k \neq i} \exp(-|| x_i - x_k ||^2 / 2\sigma_i^2)}
\end{equation}

Here, data points $x_j$ that are close to $x_i$ yield a high probability, while widely separated data points result in nearly infinitesimal values. The Gaussian distributions are centred at $x_i$  with a standard deviation $\sigma_i$. Optimal standard deviation values are determined by binary searching the value of $\sigma_i$ that produces a probability distribution satisfying:

\begin{equation}
 \log_2 \left( \textit{perplexity} \right) =  -\sum_j  p_{j|i} \log_2 (p_{j|i}),
\end{equation}

where \textit{perplexity} is a fixed value representing a smooth measure of the effective number of neighbours. 

The t-SNE algorithm diverges from SNE by requiring a symmetric conditional probability. Therefore, a symmetrized conditional probability is defined as:

\begin{equation}
 p_{ij} = \frac{p_{j|i} + p_{i|j}}{2N},
\end{equation}

where $N$ is the number of data points. Subsequently, a similarity probability distribution over the data points in the low-dimensional space is defined using a Student-t distribution with one degree of freedom (Cauchy distribution), instead of the Gaussian used in SNE. The joint similarity probabilities $q_{ij}$ are:

\begin{equation}
q_{ij} = \frac{(1 + || y_i - y_j ||^2)^{-1}}{\sum_{k \neq l} (1 + || y_k - y_l ||^2)^{-1}}.
\end{equation}

If the $y$ data points correctly model the similarity between the high-dimensional data points $x$, then the conditional probabilities $p_{ij}$ and $q_{ij}$ will be equal. t-SNE aims to find a low-dimensional data representation that minimizes the differences between these probabilities, using the Kullback-Leibler divergence to define the cost function:

\begin{equation}
C = KL(P \| Q) = \sum_{i}\sum_{j} p_{ij} \log \frac{p_{ij}}{q_{ij}}
\end{equation}

where $p_{ii}$ and $q_{ii}$ are set to zero. This cost function is minimized using gradient descent, where the gradient is:

\begin{equation}
\frac{\delta C}{\delta y_i } = 4 \sum_j (p_{ij} - q_{ij}) (y_i - y_j) (1 + ||y_i - y_j ||^2)^{-1}
\end{equation}

Gradient descent starts with map data points randomly selected from an origin-centered, isotropic, small-variance Gaussian, and is mathematically updated as follows:

\begin{equation}
\mathit{Y}^{(t)} = \mathit{Y}^{(t-1)} + \eta \frac{\delta C}{\delta \mathit{Y}} + \alpha(t) \left( \mathit{Y}^{(t-1)} - \mathit{Y}^{(t-2)} \right) 
\end{equation}

where $\mathit{Y}^{(t)}$ indicates the solution at iteration $t$, $\alpha(t)$ represents the momentum at iteration $t$, and $\eta$ is the learning rate. This process iterates until the specified number of iterations is reached. We implemented this algorithm using the Scikit-Learn Python package \citep{scikit-learn} with a learning rate of 500 and a low-dimensional space of 2. The other hyperparameters, such as perplexity (30), iterations (1000), and early exaggeration (12), are left at their default values.

\textbf{Uniform Manifold Approximation and Projection (UMAP)} is a computational technique designed by \cite{McInnes2018}. Similar to t-SNE, UMAP forms probability distributions over the data points of high-dimensional data. This is first achieved with the following conditional probability:

\begin{equation}
v_{j|i} = \exp \left( - \frac{d(x_i, x_j) - \rho_i}{\sigma_i} \right),
\end{equation}

where $d(x_i, x_j)$ is the distance between $x_i$ and $x_j$ in a pre-chosen metric space (e.g., Euclidean, Chebyshev, or Minkowski), $\rho_i$ is the distance to the nearest neighbour of $x_i$, and $\sigma_i$ is a normalizing factor determined by binary searching for a value that satisfies:

\begin{equation}
 \log_2 \left(  n \right) =  \sum_j \exp \left(  - \frac{\text{knn-dists}(x_i, x_j) - \rho_{\text{fnn}}}{\sigma_i} \right).
\end{equation}

Here, $n$ is the number of neighbors to consider, $\text{knn-dists}(x_i, x_j)$ represents the distance to the $k$-nearest-neighbor (a method used to identify the $k$ closest data points based on a specified distance metric), approximated using the Nearest-Neighbor-Descent (NND) algorithm \citep{Dong2011}, and $\rho_{\text{fnn}}$ is the distance to the first nearest neighbor of the NND algorithm.

The symmetrization of the probabilities is achieved through triangular conormalization of the fuzzy logic, expressed as:

\begin{equation}
v_{ij} = v_{j|i} + v_{i|j} - v_{j|i} \cdot v_{i|j}
\end{equation}

Subsequently, a similarity probability distribution over the low-dimensional data points is defined as:

\begin{equation}
w_{ij} = \left( 1 + a ||y_i - y_j ||^b \right)^{-1}
\end{equation}

where $y$ are the low-dimensional data points, and $a$ and $b$ are positive constants chosen by non-linear least squares fitting $f(z) = ( 1+ a z^{2b} )^{-1}$ against the curve:

\begin{equation}
g(z)=
\begin{cases}
1 & \text{if } z \leq \textit{min\_dist} \\
\exp[-(z -\textit{min\_dist})] & \text{otherwise}
\end{cases}
\end{equation}

Here, \textit{min\_dist} is the minimal distance ensuring that low-dimensional points are packed together, $z$ ranges from 0 to 1, and can be adjusted based on the desired clustering tightness. To find the low-dimensional data representation minimizing the differences between $v_{ij}$ and $w_{ij}$, the following cost function is defined:

\begin{equation}
  C = \sum_{i \neq j} \left[ v_{ij} \log \left(\frac{v_{ij}}{w_{ij}} \right) + (1 - v_{ij}) \log \left( \frac{1 - v_{ij}}{1 - w_{ij}} \right) \right]
\end{equation}

where $v_{ji}$ is calculated only for the $n$ approximate nearest neighbors and $v_{ji} = 0$ for all other $j$. The gradient of this cost function is:

\begin{equation}
\begin{split}
\frac{\delta C}{\delta y_i } = & \sum_j  \frac{-2ab ||y_i -y_j||^{2(b-1)}} {1+||y_i -y_j||^2} v_{ij}(y_i -y_j) + \\
& \sum_j \frac{2b\, v_{ij}\,(y_i -y_j)}{(0.001+||y_i -y_j||^2)(1+a||y_i -y_j||^{2b})}
\end{split}
\end{equation}

The first term represents an attractive force between vertices $i$ and $j$ at coordinates $y_i$ and $y_j$, respectively, while the second term represents a repulsive force.

The algorithm initializes the low-dimensional points with a spectral embedding of the fuzzy 1-skeleton graph using NND \footnote{For a more detailed explanation, see Section 4 of \cite{McInnes2018}}. To minimize the cost function, the algorithm uses gradient descent, updating as follows:

\begin{equation}
\mathit{Y}^{(t)} = \mathit{Y}^{(t-1)} + \beta(t) \frac{\delta C}{\delta \mathit{Y}} 
\end{equation}

where $\beta(t) = 1.0 - t / T$ is a variable learning rate, and $T$ is the number of epochs. To implement this technique, we use the UMAP-learn Python package \footnote{ \url{https://umap-learn.readthedocs.io/en/latest/index.html}}, specifying the number of neighbours $n$ as 100, dimensionality space as 2, and leaving the default hyperparameters for \textit{min\_dist} (0.1) and the number of epochs (200).

In Figure \ref{fig:tSNE_UMAP}, we present the dimensionality reduction results of the last convolutional layer of the U-Net \gizmo{}+\gadget{} for the SZ predictions. Consistent results are obtained with the corresponding X-ray model. The top row of the figure displays the outcomes from applying t-SNE, while the bottom row shows the results using UMAP. Each row is divided into three distinct graphs corresponding to different mass intervals: $13.5 \leq \log_{10} (M_{200}\, [h^{-1} M_{\odot}]) < 14.05$; $14.05 \leq \log_{10} (M_{200}\, [h^{-1} M_{\odot}]) < 14.65$; and $14.65 \leq \log_{10} (M_{200}\, [h^{-1} M_{\odot}]) < 15.5$.  Since the bi-dimensional data points can be very close to each other or even overlap, we decided to visualize the data points by creating smooth Gaussian density estimations for each dataset (\gizmo{}\_\textsc{hy}, \gadget{}\_\textsc{hy}, \gadget{}\_\textsc{dm}, and \textsc{Music\_dm}). Then for better visualisation, we chose to display a contour only for the values that have a density above the mean plus one standard deviation. Upon analyzing the figure, we observed that the data points related to \gadget{}\_\textsc{hy}, \gadget{}\_\textsc{dm}, and \textsc{Music\_dm} are situated close to each other, often showing similar spatial distributions. In all cases, the major outliers are the \gizmo{}\_\textsc{hy} data points. This indicates that the U-Net has learned to differentiate properties of the simulations, demonstrating that the \gadget{}\_\textsc{dm} and \textsc{Music\_dm} simulations have properties similar to the \gadget{}\_\textsc{hy} simulation on which the U-Net was trained. This observation strongly aligns with the previous results noted in Section \ref{sec-4.2} and Appendix \ref{Apx:DMonly_MUSIC}, where observable predictions of the U-Net \gizmo{}+\gadget{} on the DM-only simulations used in this study show similarities to the observables of the \gadget{} simulation. Apart from these findings, we can only speculate on the meaning of the data point distributions within each mass interval. The t-SNE results show that the contours intersect more with the \gizmo{} data points as the cluster masses increase, which may indicate that the learned properties between simulations coincide more at higher masses. This is in agreement with the fact that the scaling relations are self-similar between simulations for massive clusters, being astrophysical effects less relevant \citep{Cui2022}. However, this trend is not observed in the UMAP results, potentially due to the algorithmic and probabilistic differences between t-SNE and UMAP. A plausible physical explanation for the differences between \gizmo{}\_\textsc{hy} runs and the others could be due to its strong feedback, which alters the internal structure of the mass distribution, specially the gas component.

\section{Conclusions}\label{sec-5}

In this paper, we present a novel approach for simulating multiview maps of galaxy clusters, specifically SZ (Sunyaev-Zel’dovich) and X-ray bolometric maps, from dark matter-only simulations aiming at reducing computational costs mimicking the results of hydrodynamical simulations. The deep learning architecture employed in this work is based on the U-Net model. Our primary datasets consist of both hydrodynamical and DM-only simulations from \thethree{} project. Various normalization techniques were tested to mitigate the impact of scale differences, with \textit{Standardisation for Log10} showing the best performance (Appendix \ref{Apx:training_metrics}). To evaluate the effectiveness of the models, we employ both the test datasets derived from the hydrodynamical simulations and the datasets from DM-only simulations. We utilize three different metrics to assess model performance. Additionally, we compare the observable-mass  scaling relations of the generated maps with those of the ground-truth datasets. Our main results are as follows:

\begin{enumerate}
    \item The metrics applied to the test dataset showed that based on the Maximum Mean Discrepancy (MMD) in Figure \ref{fig:Test_MMD}, the SZ predictions exhibited, for the most part, discrepancies less than 20\% map noise and decreasing to 5\% in the case of massive clusters. For X-ray predictions, discrepancies were generally lower than 10\% of noise, indicating higher statistical similarity. Figure \ref{fig:Test_MRD} shows that the MRD predictions for both observables tended to overestimate low-mass clusters and underestimate higher masses. U-Nets using \textsc{Gadget} inputs show less scatter and thus more accurate predictions. Figure \ref{fig:Test_SSIM} reveals that X-ray map predictions demonstrated better structural similarity (SSIM) compared to SZ maps, especially for massive clusters. The U-Net \gizmo{}+\gadget{} model effectively predicted the scaling relations for both SZ and X-ray observables, as depicted in Figures \ref{fig:SZ_scaling_train} and \ref{fig:Xray_scaling_train}. Notably, the model exhibited deviations in predictions for low-mass clusters. The scaling parameters  (equations \ref{eq:SZ-fit} and \ref{eq:Xray-fit}) demonstrated better alignment with the \gadget{} simulation, showing percentage errors averaging (0.8±0.3)\%, compared to \gizmo{}, which presented percentage errors averaging (2.4±0.4)\%.
    
    \item Using total mass maps from the DM-only simulations as input, the U-Net \textsc{Gadget}+\textsc{Gizmo} model predicted observables similar to the \textsc{Gadget} simulation across all cluster masses (Figure \ref{fig:MetricsComparison}), with X-ray predictions being more accurate than SZ, which is consistent with the testing results. The scaling relations (Figures \ref{fig:SZ_FullDMonly_M200} and \ref{fig:Xray_FullDMonly_M200}) indicated that before the  pivotal point around $2 \times 10^{14}h^{-1}M_{\odot}$ the predictions deviated from the ground-truth datasets. Nonetheless, for clusters with masses above this pivot point, the models were more precise with percentage errors averaging (0.5±0.1)\%.
    

    \item Analysis of the gas mass profiles and dimensionality reduction techniques indicated that the U-Net models learned to differentiate between simulations, particularly distinguishing \gizmo{} from others. In general, for generating a \gizmo{}-like catalogue the model has to be trained only with \gizmo{} data. On the contrary, if the model is trained with a combined dataset, the model will provide \gadget{} properties. 
\end{enumerate}

In addition to the results presented, we simulated several regions using the \gizmo{} gravity solver, where we found that the mass profiles of the clusters and the U-Nets' predicted observables were consistent with those obtained from the \gadget{}-X DM-only simulation. This indicates that the differences between \gizmo{} and \gadget{} primarily stem from hydrodynamics and subgrid physics.

The findings highlight the potential of U-Net models in predicting observable maps of galaxy clusters. Nonetheless, we identify similar constraints to those noted by \cite{Chadayammuri2023} in a comparable study on the  {\sc ILLUSTRIS TNG300} simulation. A similar problem we faced in our study is that matching halos between hydrodynamical and DM-only simulations with the "same" initial conditions is not trivial due to the chaotic nature of N-body simulations.

Training with multiple simulations offers two distinct perspectives. When predicting from simulations that were included in the training set, the algorithm can accurately differentiate between simulations and predict observable maps from the total mass with high precision. Conversely, when predicting from unseen simulations, such as those with DM-only, the algorithm makes predictions based on the learned properties from the training data, resulting in lower accuracy for low-mass halos, as illustrated in Figure \ref{fig:Xray_FullDMonly_M200}.  Generally, the models are capable of generating simulated catalogues with high fidelity for DM-only halos with masses greater than $2 \times 10^{14} h^{-1} M_{\odot}$. For smaller halos, however, the influence of astrophysical effects such as star formation, supernova feedback, and gas cooling becomes significant, introducing complexities that these models may not fully capture.

Dimensionality reduction methods such as t-SNE and UMAP can aid in understanding ML models superficially, but further investigation is needed to relate the techniques' data point characteristics to model inputs and outputs. In addition, by considering Domain Adaptation techniques \citep{Farahani2020:DA} models could learn only common features of \gizmo{} and \gadget{}, rendering the models more independent across multiple simulations.  This will be considered in a forthcoming study. 


This study demonstrates that deep learning is a promising method for improving the mapping between mass and observables using idealized simulations. Although challenges persist in the method's accuracy—particularly for low-mass clusters—and in its interpretation, the findings provide a valuable foundation for future research aimed at efficiently reproducing observable features. Incorporating a broader range of simulations, as done by \cite{CAMELS}, could further enhance our understanding of the relationship between mass and baryons. Furthermore, as highlighted by \cite{deAndres2024}, the potential benefits of utilizing adversarial deep learning models, such as the Wasserstein Generative Adversarial Network \citep[WGAN,][]{WGAN}, may yield comparable results. 

The future applications of the trained models, which are left for a forthcoming work, are manifold:
\begin{enumerate}
    \item Exploring the model's performance across different cosmologies.
    \item The models can be used to populate Gpc-size dark-matter-only simulations with ICM observational maps, thereby increasing the statistical sample of galaxy-cluster-sized objects with available ICM information.
    \item  By extracting many different light-cones \citep{Zandanel2018} from this new large computational volumes\footnote{for instance, UNITSIMS, \url{http://www.unitsims.org}}, the cosmic variance of full-sky surveys, such as eROSITA \citep{Liu2022}, can be studied.
    \item We demonstrated that the generated 2D observational maps not only reproduce  the expected scaling laws but also reveal detailed structural features. These features enable the analysis of the dynamical state \citep{DeLuca2021} of new large-volume samples from X-ray and SZ AI-assisted simulations.

\end{enumerate}

\section*{Acknowledgements}

We would like to express our gratitude to the anonymous referees for the comments that greatly helped us improve the quality of the manuscript. 
The authors acknowledge The Red Española de Supercomputación for granting computing time for running most of the hydrodynamical simulations of \thethree{} galaxy cluster project in the Marenostrum supercomputer at the Barcelona Supercomputing Center. 
D.d.A., W.C. and G.Y. would like to thank Ministerio de Ciencia e Innovación for financial support under project grant PID2021-122603NB-C21. WC is supported by the STFC AGP Grant ST/V000594/1 and the Atracci\'{o}n de Talento Contract no. 2020-T1/TIC-19882 granted by the Comunidad de Madrid in Spain. He also thanks the ERC: HORIZON-TMA-MSCA-SE for supporting the LACEGAL-III project with grant number 101086388 and the China Manned Space Project for its research grants. M.D.P. and A.F. acknowledge financial support from PRIN 2022 (Mass and selection biases of galaxy clusters: a multi-probe approach - n. 20228B938N) and from Sapienza Universita
di Roma, thanks to Progetti di Ricerca Medi 2022, RM1221816758ED4E.


\section*{Data Availability}
 
The results shown in this work use data from The {\sc Three Hundred} galaxy clusters sample. The data is freely available upon request following the guidelines of The Three Hundred collaboration, at  \url{https://www.the300-project.org}.



\bibliographystyle{rasti}
\bibliography{main} 




\appendix

\section{Training Normalization Metrics}
\label{Apx:training_metrics}

\begin{figure*}
\includegraphics[width=0.7\textwidth]{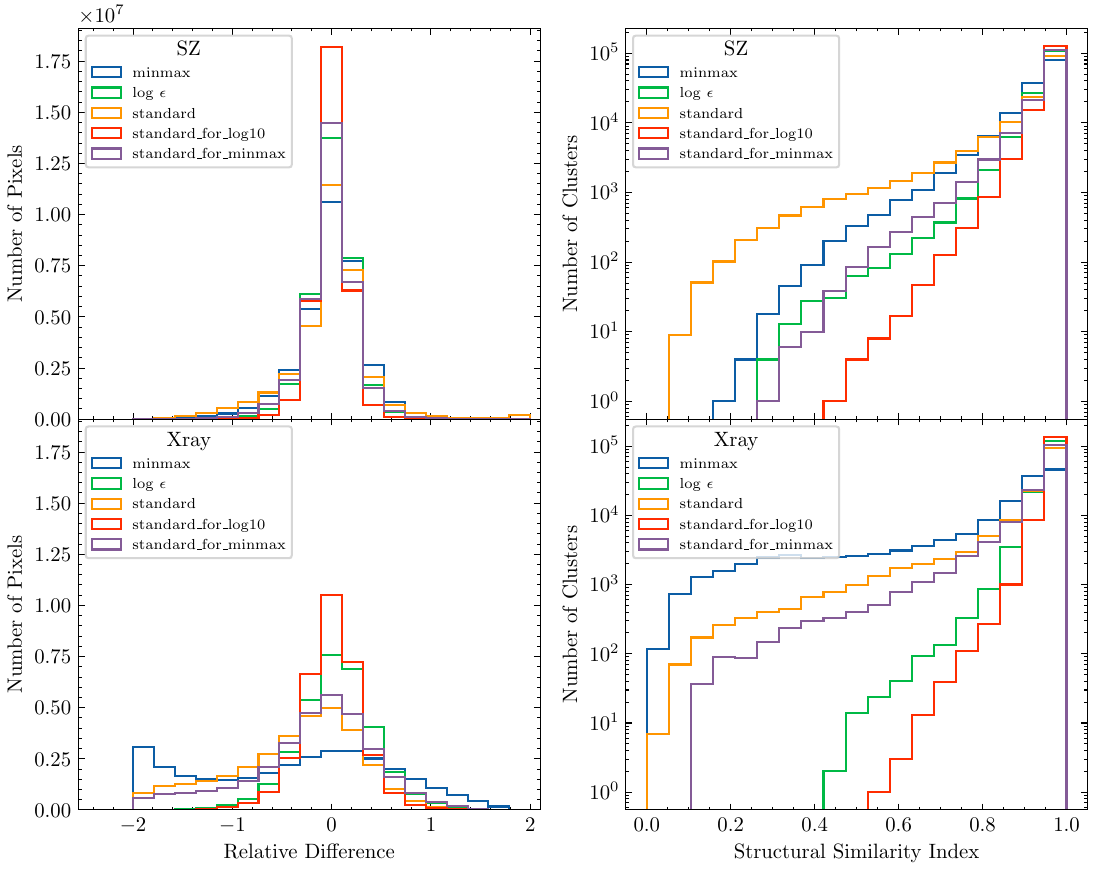}
\caption{ Histograms depicting the Relative Difference of each pixel (left) and the Structural Similarity Index Measure (SSIM, right) of clusters for five distinct normalizations applied to the training data (Appendix \ref{Apx:training_metrics}). These metrics assess the discrepancies between the ground-truth observable maps and the observable predictions of the U-Net \gizmo{}+\gadget{}. }
 \label{fig:Normalization_Metrics}
\end{figure*}

Before training all the models, we tested five types of normalizations for each map $x$ to determine the most suitable for our data $X$. These normalizations are:

\begin{itemize} 

\item \textbf{Min-Max:} This scales the data to a specified range, typically between 0 and 1, while preserving the relative relationships between data points. The equation for min-max normalization is as follows:

\begin{equation*}
    x' = \frac{{x - \min(X)}}{{\max(X) - \min(X)}}
    \label{eq:minmax}
\end{equation*}

Here, $\min(X)$ and $\max(X)$ denote the minimum and maximum values in the dataset.

\item \textbf{Log {\boldmath $\epsilon$}:} This applies a parametric logarithmic function to the data \citep[from, ][]{Rothschild2022}, that helps mitigate the influence of extremely low values and compresses the range of the data. It is defined as:

\begin{equation*}
    x' = \log_{\frac{1}{\epsilon}} \left( \frac{x + \epsilon}{\epsilon} \right) 
    \label{eq:logepsilon}
\end{equation*}

In this equation, $\epsilon$ is a small positive constant chosen as the $10^{th}$ percentile of the dataset.

\item \textbf{Standard:} Also known as Z-score normalization, this transforms the data to have zero mean and unit variance, effectively centring the data around the mean and adjusting the scale to account for variations. The equation for standardization is given by:

\begin{equation*}
    x' = \frac{{x - \mu(X)}}{{\sigma(X)}} 
    \label{eq:standard}
\end{equation*}

where $\mu$ is the mean and $\sigma$ is the standard deviation of the whole dataset.

\item \textbf{Standardization for Log10:} This normalization technique involves applying standard normalization specifically to a logarithmically transformed dataset. It accounts for the dynamic range of the data while maintaining most of the statistical properties. The process involves two steps: first, changing the scale of the dataset by $X_{\text{log10}} = \log_{10} (X + \min(X))$, and then using the standard normalization on $X_{\text{log10}}$, as follows:

\begin{equation*}
    x' = \frac{x_{\text{log10}} - \mu(X_{\text{log10}})}{\sigma(X_{\text{log10}})}
    \label{eq:standardforlog10}
\end{equation*}

Here, $X_{\text{log10}}$ is the dataset after the logarithmic transformation, $x_{\text{log10}}$ is a map from $X_{\text{log10}}$.

\item \textbf{Standardization for Min-Max:} This normalization combines standard normalization and min-max normalization. The process involves two steps: first, the min-max transformation is applied as $X_{\text{minmax}} = (X - \min(X)) / (\max(X) - \min(X))$, then the standard normalization is performed on $X_{\text{minmax}}$ as follows:

\begin{equation*}
    x' = \frac{x_{\text{minmax}} - \mu(X_{\text{minmax}})}{\sigma(X_{\text{minmax}})}
    \label{eq:standardforminmax}
\end{equation*}

Here, $X_{\text{minmax}}$ is the dataset after the min-max transformation, $x_{\text{minmax}}$ is a map from $X_{\text{minmax}}$.

\end{itemize}

In Figure \ref{fig:Normalization_Metrics}, we present the histograms of the relative difference of each pixel ($2 (x_i - y_i) / (x_i + y_i)$) and the SSIM (equation (\ref{eq:SSIM})) of each cluster for the normalizations used in training the U-Net \gizmo{}+\gadget{} model. The figure reveals that \textit{Standardization for Log10} yields the best results, as most data points are close to 0 in the relative difference case and closer to 1 in the SSIM case. Overall, the most suitable normalizations for this type of data are those that consider the data range and the variation of random pixels.

\section{Scaling Parameters}
\label{Apx:parameter_tables}

Before to estimate the parameters of the fit, we first transform the power-law relationships from equations (\ref{eq:SZ-fit}) and (\ref{eq:Xray-fit}) into:

\begin{equation} 
\log_{10} \left( E(z)^{-2/3} d_A^2 Y \right)  = {B_{SZ}} \log_{10} \left( M \right ) + {A_{SZ}}, 
\end{equation}

and

\begin{equation} 
\log_{10} \left( E(z)^{-7/3} L_X  \right)  = {B_{X}} \log_{10} \left( M \right ) + {A_{X}}, 
\end{equation}

These equations are then linearly fitted using the least squares method\footnote{The fitting was done with the Numpy's library function \texttt{numpy.polyfit}, for more information we refer to \url{https://numpy.org/doc/stable/}}. Since the data shows a pivot point at $2\times 10^{14} h^{-1} M_\odot$, we fit the scaling relations as a broken power-law. The parameters with subscripts $SZ_1$ and $X_1$ represent the values before the pivot point, while $SZ_2$ and $X_2$ represent the values after the pivot point.

In Tables \ref{tab:SZ_training} and \ref{tab:Xray_training}, we present the fitted scaling parameters for the predicted SZ and X-ray observables when using the U-Net \gizmo{}+\gadget{} with the test dataset. The percentage error is calculated as $100 \times |A_{\text{pred}}-A_{\text{true}}|/|A_{\text{true}}|$ compared to the ground-truth parameters and is displayed next to the prediction parameters.

Tables \ref{tab:SZ_DMonly} and \ref{tab:Xray_DMonly} show the fitted scaling parameters for the predicted SZ and X-ray observables for all U-Nets when using the \gadget{}-X DM-only dataset. For the U-Net \gizmo{}+\gadget{}, the percentage errors are calculated against the ground-truth dataset that was closest to its trend, which in most cases was the \gadget{} simulation. Similarly, in Tables \ref{tab:SZ_MUSIC} and \ref{tab:Xray_MUSIC}, we present the fitted scaling parameters for the predicted SZ and X-ray observables for all U-Nets when using the \music{} DM-only dataset.

\begin{table*}
\centering
\resizebox{\textwidth}{!}{%
\begin{tabular}{lllllllll}
\cline{2-9}
& $A_{SZ_1}$    & \% error    & $B_{SZ_1}$    & \% error    & $A_{SZ_2}$    & \% error    & $B_{SZ_2}$    & \% error    \\ \hline
True \gizmo{}        & -35.25 ± 0.08 & -          & 2.102 ± 0.006 & -          & -28.87 ± 0.01 & -          & 1.656 ± 0.001 & -          \\
Prediction \gizmo{}  & -34.64 ± 0.16 & 1.7 ± 0.5 & 2.058 ± 0.012 & 2.1 ± 0.6 & -28.75 ± 0.03 & 0.4 ± 0.1 & 1.647 ± 0.002 & 0.5 ± 0.1 \\
True \gadget{}       & -29.04 ± 0.05 & -          & 1.670 ± 0.003 & -          & -27.84 ± 0.01 & -          & 1.588 ± 0.001 & -          \\
Prediction \gadget{} & -28.91 ± 0.10 & 0.4 ± 0.4 & 1.660 ± 0.007 & 0.6 ± 0.5 & -27.80 ± 0.03 & 0.1 ± 0.1 & 1.584 ± 0.002 & 0.2 ± 0.1 \\ \hline
\end{tabular}%
}
\caption{Fitted scaling law parameters for the predicted SZ observables using the U-Net \gizmo{}+\gadget{} with the test dataset (Section \ref{sec-2.2}). The percentage errors are calculated against the ground-truth data as $100 \times |A_{\text{pred}}-A_{\text{true}}|/|A_{\text{true}}|$. Since there is a pivot point in steepness at roughly $2\times 10^{14} h^{-1} M_\odot$, the $SZ_1$ subscripts reflect the fitted results for masses below this threshold, while the $SZ_2$ subscripts represent the fitted findings for masses above it. }
\label{tab:SZ_training}
\end{table*}

\begin{table*}
\centering
\resizebox{\textwidth}{!}{%
\begin{tabular}{lllllllll}
\cline{2-9}
& $A_{X_1}$    & \% error    & $B_{X_1}$     & \% error    & $A_{X_2}$    & \% error   & $B_{X_2}$     & \% error    \\ \hline
True \gizmo{}        & 9.19 ± 0.11  & -          & 2.391 ± 0.008 & -          & 18.23 ± 0.03 & -         & 1.765 ± 0.002 & -          \\
Prediction \gizmo{}  & 8.35 ± 0.22  & 9.1 ± 2.7 & 2.447 ± 0.015 & 2.3 ± 0.7  & 18.51 ± 0.06 & 1.6 ± 0.3 & 1.742 ± 0.004 & 1.3 ± 0.2 \\
True \gadget{}       & 14.24 ± 0.11 & -          & 2.078 ± 0.008 & -          & 25.14 ± 0.05 & -         & 1.322 ± 0.003 & -          \\
Prediction \gadget{} & 14.64 ± 0.21 & 2.8 ± 1.6  & 2.045 ± 0.015 & 1.6 ± 0.8 & 25.24 ± 0.09 & 0.4 ± 0.4 & 1.310 ± 0.006 & 0.9 ± 0.5 \\ \hline
\end{tabular}%
}
\caption{Fitted scaling law parameters for the predicted X-ray observables using the U-Net \gizmo{}+\gadget{} with the test dataset (Section \ref{sec-2.2}). The percentage errors are calculated against the ground-truth data as $100 \times |A_{\text{pred}}-A_{\text{true}}|/|A_{\text{true}}|$. Since there is a pivot point in steepness at roughly $2\times 10^{14} h^{-1} M_\odot$, the $X_1$ subscripts reflect the fitted results for masses below this threshold, while the $X_2$ subscripts represent the fitted findings for masses above it. }
\label{tab:Xray_training}
\end{table*}

\begin{table*}
\centering
\resizebox{\textwidth}{!}{%
\begin{tabular}{lllllllll}
\cline{2-9}
& $A_{SZ_1}$    & \% error    & $B_{SZ_1}$    & \% error    & $A_{SZ_2}$    & \% error    & $B_{SZ_2}$    & \% error    \\ \hline
True \gizmo{}                       & -35.25 ± 0.08 & -          & 2.102 ± 0.006 & -          & -28.87 ± 0.01 & -          & 1.656 ± 0.001 & -          \\
Prediction U-Net \gizmo{}           & -38.79 ± 0.33 & 10.1 ± 1.0 & 2.348 ± 0.023 & 11.7 ± 1.1 & -28.79 ± 0.04 & 0.3 ± 0.2 & 1.650 ± 0.003 & 0.3 ± 0.2 \\
True \gadget{}                      & -29.04 ± 0.05 & -          & 1.670 ± 0.003 & -          & -27.84 ± 0.01 & -          & 1.588 ± 0.001 & -          \\
Prediction U-Net \gadget{}          & -29.41 ± 0.15 & 1.3 ± 0.5  & 1.695 ± 0.011 & 1.5 ± 0.7  & -27.83 ± 0.04 & 0.1 ± 0.1 & 1.585 ± 0.002 & 0.2 ± 0.2 \\
Prediction U-Net \gizmo{}+\gadget{} & -39.47 ± 0.30 & 12.0 ± 0.9* & 2.404 ± 0.022 & 14.4 ± 1.1* & -27.97 ± 0.04 & 0.5 ± 0.1  & 1.595 ± 0.003 & 0.4 ± 0.2  \\ \hline
\end{tabular}%
}
\caption{Fitted scaling law parameters for the predicted SZ observables for all U-Nets using the \gadget{}-X DM-only dataset (Section \ref{sec-2.3}). The percentage errors are calculated against the ground-truth data. For U-Net \gizmo{} and U-Net \gadget{}, these errors are computed using their respective ground-truth datasets. For U-Net \gizmo{}+\gadget{}, the percentage errors are calculated against the ground-truth dataset that most closely aligns with its trend, primarily the \gadget{} simulation. The percentage errors marked with an asterisk (*) are the only ones for U-Net \gizmo{}+\gadget{} compared with \gizmo{}. Since there is a pivot point in steepness at roughly $2\times 10^{14} h^{-1} M_\odot$, the $SZ_1$ subscripts reflect the fitted results for masses below this threshold, while the $SZ_2$ subscripts represent the fitted findings for masses above it. }
\label{tab:SZ_DMonly}
\end{table*}

\begin{table*}
\centering
\resizebox{\textwidth}{!}{%
\begin{tabular}{lllllllll}
\cline{2-9}
& $A_{X_1}$    & \% error      & $B_{X_1}$     & \% error    & $A_{X_2}$    & \% error    & $B_{X_2}$     & \% error    \\ \hline
True \gizmo{}                       & 9.19 ± 0.11  & -            & 2.391 ± 0.008 & -          & 18.23 ± 0.03 & -          & 1.765 ± 0.002 & -          \\
Prediction U-Net \gizmo{}           & -1.09 ± 0.44 & 111.9 ± 5.2 & 3.107 ± 0.032 & 29.9 ± 1.4 & 18.63 ± 0.08 & 2.2 ± 0.5  & 1.739 ± 0.005 & 1.4 ± 0.3 \\
True \gadget{}                      & 14.24 ± 0.11 & -            & 2.078 ± 0.008 & -          & 25.14 ± 0.05 & -          & 1.322 ± 0.003 & -          \\
Prediction U-Net \gadget{}          & 15.48 ± 0.32 & 8.7 ± 2.4    & 1.986 ± 0.023 & 4.4 ± 1.2 & 25.02 ± 0.11 & 0.5 ± 0.5 & 1.323 ± 0.007 & 0.1 ± 0.6  \\
Prediction U-Net \gizmo{}+\gadget{} & 1.70 ± 0.53  & 88.1 ± 3.9  & 2.961 ± 0.038 & 42.5 ± 1.9 & 25.13 ± 0.11 & 0.0 ± 0.5  & 1.318 ± 0.007 & 0.3 ± 0.6 \\ \hline
\end{tabular}%
}
\caption{Fitted scaling law parameters for the predicted X-ray observables for all U-Nets using the \gadget{}-X DM-only dataset (Section \ref{sec-2.3}). The percentage errors are calculated against the ground-truth data. For U-Net \gizmo{} and U-Net \gadget{}, these errors are computed using their respective ground-truth datasets. For U-Net \gizmo{}+\gadget{}, the percentage errors are calculated against the ground-truth dataset that most closely aligns with its trend, the \gadget{} simulation. Since there is a pivot point in steepness at roughly $2\times 10^{14} h^{-1} M_\odot$, the $X_1$ subscripts reflect the fitted results for masses below this threshold, while the $X_2$ subscripts represent the fitted findings for masses above it. }
\label{tab:Xray_DMonly}
\end{table*}

\begin{table*}
\centering
\resizebox{\textwidth}{!}{%
\begin{tabular}{lllllllll}
\cline{2-9}
& $A_{SZ_1}$    & \% error    & $B_{SZ_1}$    & \% error    & $A_{SZ_2}$    & \% error    & $B_{SZ_2}$    & \% error    \\ \hline
True \gizmo{}              & -35.25 ± 0.08 & -          & 2.102 ± 0.006 & -          & -28.87 ± 0.01 & -          & 1.656 ± 0.001 & -          \\
Prediction U-Net \gizmo{}  & -33.91 ± 0.43 & 3.8 ± 1.2 & 2.014 ± 0.031 & 4.2 ± 1.5 & -27.89 ± 0.08 & 3.4 ± 0.3 & 1.592 ± 0.006 & 3.9 ± 0.3 \\
True \gadget{}             & -29.04 ± 0.05 & -          & 1.670 ± 0.003 & -          & -27.84 ± 0.01 & -          & 1.588 ± 0.001 & -          \\
Prediction U-Net \gadget{} & -28.72 ± 0.23 & 1.1 ± 0.8 & 1.649 ± 0.017 & 1.3 ± 1.0 & -27.52 ± 0.07 & 1.2 ± 0.3 & 1.567 ± 0.005 & 1.3 ± 0.3 \\
Prediction U-Net \gizmo{}+\gadget{} & -29.00 ± 0.23 & 0.1 ± 0.8 & 1.669 ± 0.016 & 0.1 ± 1.0 & -27.41 ± 0.07 & 1.5 ± 0.3 & 1.560 ± 0.005 & 1.8 ± 0.3 \\ \hline
\end{tabular}%
}
\caption{Fitted scaling law parameters for the predicted SZ observables for all U-Nets using the \music{} DM-only dataset (Section \ref{sec-2.3}). The percentage errors are calculated against the ground-truth data. For U-Net \gizmo{} and U-Net \gadget{}, these errors are computed using their respective ground-truth datasets. For U-Net \gizmo{}+\gadget{}, the percentage errors are calculated against the ground-truth dataset that most closely aligns with its trend, the \gadget{} simulation. Since there is a pivot point in steepness at roughly $2\times 10^{14} h^{-1} M_\odot$, the $SZ_1$ subscripts reflect the fitted results for masses below this threshold, while the $SZ_2$ subscripts represent the fitted findings for masses above it. }
\label{tab:SZ_MUSIC}
\end{table*}

\begin{table*}
\centering
\resizebox{\textwidth}{!}{%
\begin{tabular}{lllllllll}
\cline{2-9}
 & $A_{X_1}$    & \% error     & $B_{X_1}$     & \% error    & $A_{X_2}$    & \% error    & $B_{X_2}$     & \% error    \\ \hline
True \gizmo{}              & 9.19 ± 0.11  & -           & 2.391 ± 0.008 & -          & 18.23 ± 0.03 & -          & 1.765 ± 0.002 & -          \\
Prediction U-Net \gizmo{}  & 2.85 ± 0.57  & 68.9 ± 6.4 & 2.851 ± 0.041 & 19.2 ± 1.7 & 19.22 ± 0.15 & 5.5 ± 0.9  & 1.704 ± 0.010 & 3.5 ± 0.6 \\
True \gadget{}             & 14.24 ± 0.11 & -           & 2.078 ± 0.008 & -          & 25.14 ± 0.05 & -          & 1.322 ± 0.003 & -          \\
Prediction U-Net \gadget{} & 16.25 ± 0.47 & 14.2 ± 3.4  & 1.937 ± 0.034 & 6.8 ± 1.7 & 25.42 ± 0.24 & 1.1 ± 1.0  & 1.300 ± 0.016 & 1.6 ± 1.3 \\
Prediction U-Net \gizmo{}+\gadget{} & 16.60 ± 0.47 & 16.6 ± 3.4  & 1.915 ± 0.034 & 7.8 ± 1.7 & 24.94 ± 0.25 & 0.8 ± 1.0 & 1.336 ± 0.017 & 1.1 ± 1.3  \\ \hline
\end{tabular}%
}
\caption{Fitted scaling law parameters for the predicted X-ray observables for all U-Nets using the \music{} DM-only dataset (Section \ref{sec-2.3}). The percentage errors are calculated against the ground-truth data. For U-Net \gizmo{} and U-Net \gadget{}, these errors are computed using their respective ground-truth datasets. For U-Net \gizmo{}+\gadget{}, the percentage errors are calculated against the ground-truth dataset that most closely aligns with its trend, the \gadget{} simulation. Since there is a pivot point in steepness at roughly $2\times 10^{14} h^{-1} M_\odot$, the $X_1$ subscripts reflect the fitted results for masses below this threshold, while the $X_2$ subscripts represent the fitted findings for masses above it. }
\label{tab:Xray_MUSIC}
\end{table*}

\section{\music{} DM-only predictions}
\label{Apx:DMonly_MUSIC}

\begin{figure}
\includegraphics[width=0.9\columnwidth]{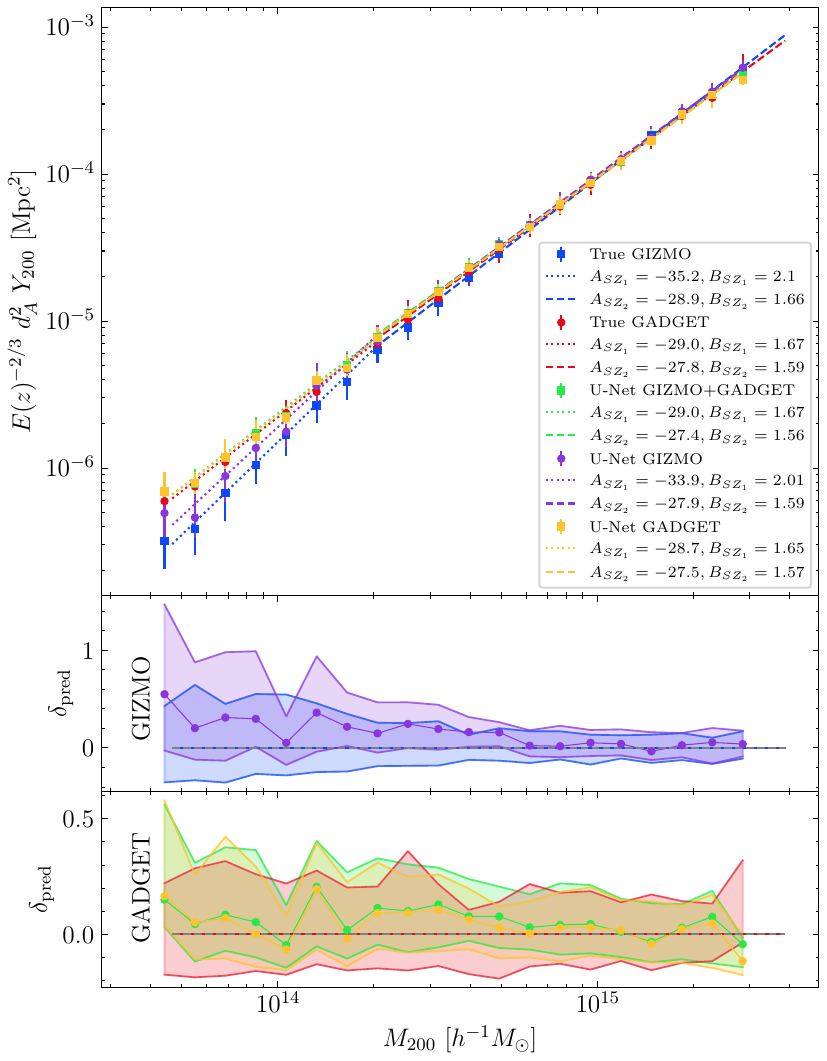}
\caption{The $Y$–$M$ scaling relation derived from the \textsc{Gizmo} (blue) and \textsc{Gadget} (red) simulations, as well as from the simulated observations generated using U-Nets: \gizmo{}+\gadget{} (green), \textsc{Gizmo} (purple), and \textsc{Gadget} (yellow). These scaling relations were obtained by using the total mass maps from the  \music{} DM-only dataset as input (Section \ref{sec-2.3}). The error bars correspond to the $16^{\text{th}}$ to $84^{\text{th}}$ percentile range. On the top graph, the lines depict the fitted results obtained from equation (\ref{eq:SZ-fit}).  Given that the pivot point in steepness occurs at approximately $2\times 10^{14} h^{-1} M_\odot$, the dotted lines and the $SZ_1$ subscripts represent the fitted results for masses below this threshold, while the dashed lines and the $SZ_2$ subscripts represent the fitted results for masses above it. The bottom panels shows the median (solid lines), the $16^{\text{th}}$  and $84^{\text{th}}$ percentiles (shaded region) of the relative error $\delta_{\text{pred}} = (Y_{\text{pred}} - Y_{\text{true}}) / Y_{\text{true}}$. For the U-Net \textsc{Gizmo}+\textsc{Gadget} relative error, we consider the \textsc{Gadget} simulation as the $Y_{\text{true}}$.}
 \label{fig:SZ_MUSIC_M200}
\end{figure}

\begin{figure}
\includegraphics[width=0.9\columnwidth]{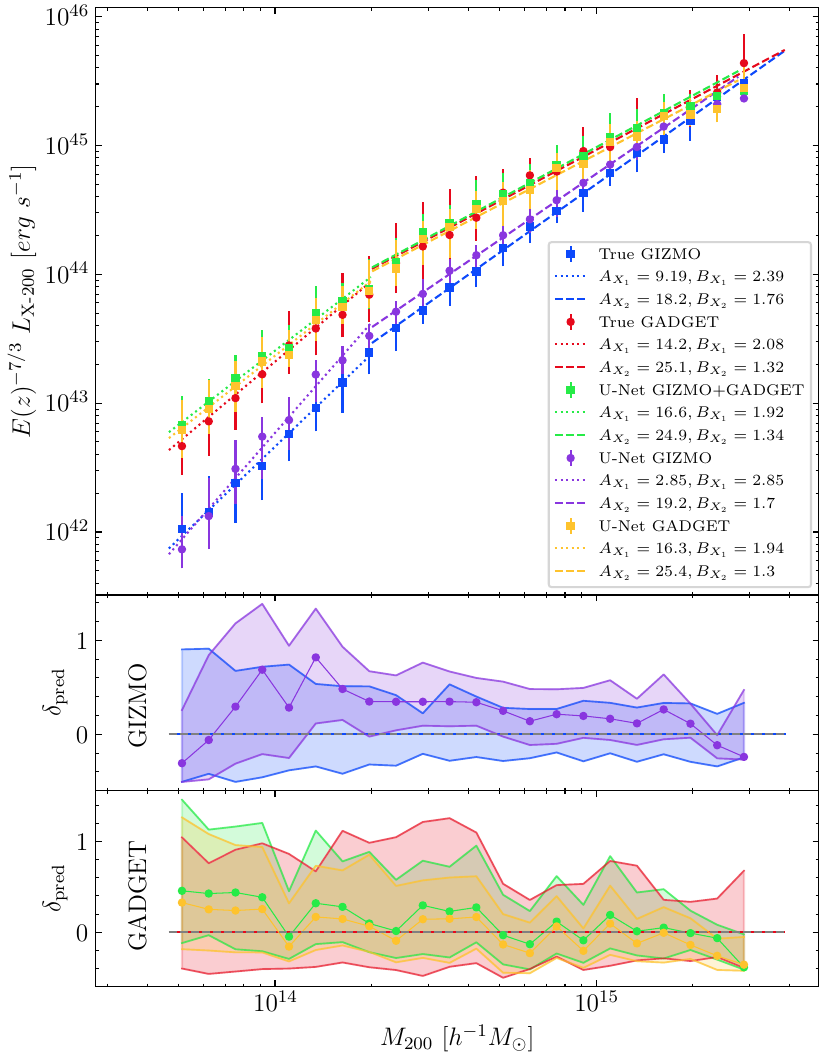}
\caption{The $L_X$–$M$ relationship derived from the \textsc{Gizmo} (blue) and \textsc{Gadget} (red) simulations, as well as from the simulated observations generated using U-Nets: \gizmo{}+\gadget{} (green), \textsc{Gizmo} (purple), and \textsc{Gadget} (yellow). These scaling relations were obtained by using the total mass maps from the \music{} DM-only dataset as input (Section \ref{sec-2.3}). The error bars correspond to the $16^{\text{th}}$ to $84^{\text{th}}$ percentile range. On the top graph, the lines are fitted according to equation (\ref{eq:Xray-fit}). Given that the pivot point in steepness occurs at approximately $2\times 10^{14} h^{-1} M_\odot$, the dotted lines and the $X_1$ subscripts represent the fitted results for masses below this threshold, while the dashed lines and the $X_2$ subscripts represent the fitted results for masses above it.  The bottom panels shows the median (solid lines), the $16^{\text{th}}$  and $84^{\text{th}}$ percentiles (shaded regions) of the relative error $\delta_{\text{pred}} = (L_{X\text{-pred}} - L_{X\text{-true}}) / L_{X\text{-true}}$ . For the U-Net \textsc{Gizmo}+\textsc{Gadget} relative error, we consider the \textsc{Gadget} simulation as the $L_{X-\text{true}}$.}
 \label{fig:Xray_MUSIC_M200}
\end{figure}

By using the \music{} DM-only dataset described in Section \ref{sec-2.3}, we plot the scaling relation of the predicted observables for all the U-Nets and calculate the percentage errors for the scaling parameters in Appendix \ref{Apx:parameter_tables}. Figure \ref{fig:SZ_MUSIC_M200} shows the median values of the cylindrical integrated $Y$ up to $R_{200}$, with error bars representing the $16^{\text{th}}$ to $84^{\text{th}}$ percentiles. The scaling relations fitted by least squares are depicted with lines, and the legend provides information on the fitting parameters. The figure indicates that the scaling of the predicted observables follows the main trend of the hydrodynamical simulations. Fitting parameters from equation (\ref{eq:SZ-fit}) demonstrate that the U-Net \gizmo{} exhibits less accuracy compared to the other models, with errors not exceeding (4.2 ± 1.5)\% before the pivot point of $2\times 10^{14} h^{-1} M_\odot$, and (3.9 ± 0.3)\% thereafter.

In Figure \ref{fig:Xray_MUSIC_M200}, the median values of the cylindrical integrated $L_X$ up to $R_{200}$ are shown, with error bars indicating the $16^{\text{th}}$ to $84^{\text{th}}$ percentiles. The least squares fitted scaling relations are represented by lines, and the parameters are detailed in the legend. Analyzing this figure, we observe that the predictions of the U-Net \gizmo{}+\gadget{} are closely related to the \gadget{} simulation scaling. This is expected as the \music{} simulations do not include radiative physics, resulting in a higher total mass distribution compared to hydrodynamical simulations. This is particularly evident in Figure \ref{fig:4sim_density_profiles}, where the median of the \music{} DM-only total mass profiles are closer to the hydrodynamical \gadget{} profiles.

\section{GIZMO and GADGET simulations Gas Profiles}
\label{Apx:Gas_profiles} 

\begin{figure*}
\includegraphics[width=1\textwidth]{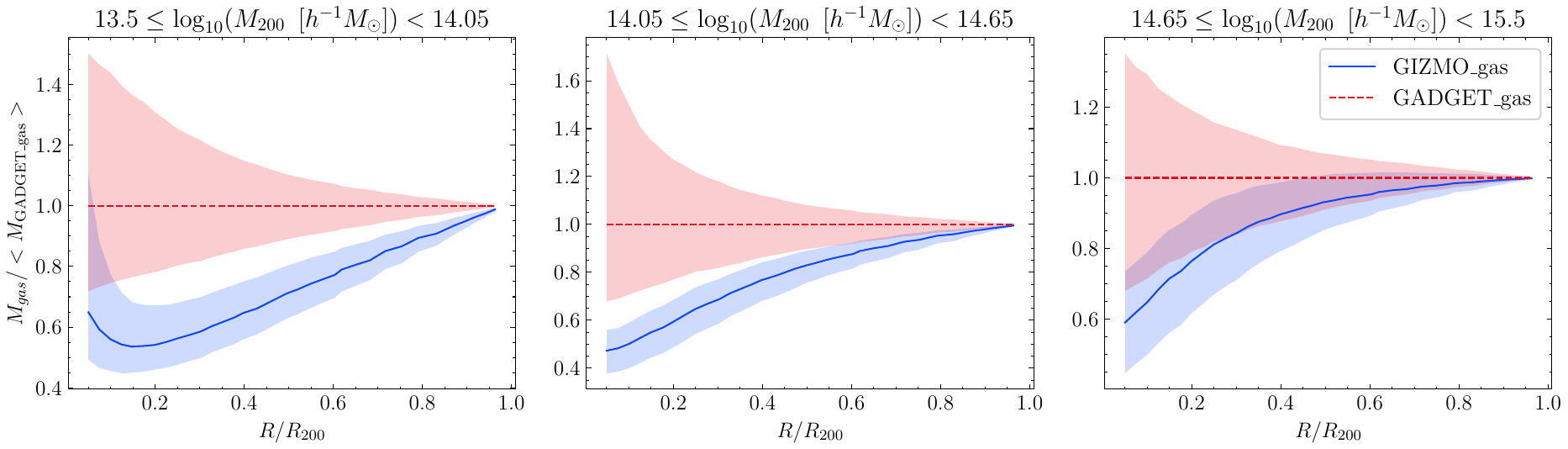}
\caption{ Gas radial profiles are depicted for the hydrodynamical simulations conducted using \gizmo{} and \gadget{}. The vertical axis indicates mass normalized by the mean radial mass of the gas in the \gadget{} dataset clusters, while the horizontal axis denotes the cluster radius normalized by their corresponding $R_{200}$. Each graph, from left to right, corresponds to the three intervals specified in the plot titles. The lines represent the median values per bin, while the shaded regions indicate the $16^{\text{th}}$ to $84^{\text{th}}$ percentiles. }
\label{fig:gas_profiles} 
\end{figure*}

We analyzed the gas mass profiles to further investigate the differences between the hydrodynamical simulations. Figure \ref{fig:gas_profiles} illustrates the gas mass profiles normalized by the \gadget{}\_gas profile values and scaled to $R/R_{200}$ across three intervals: $13.5 \leq \log_{10} (M_{200} , [h^{-1} M_{\odot}]) < 14.05$, $14.05 \leq \log_{10} (M_{200} , [h^{-1} M_{\odot}]) < 14.65$, and $14.65 \leq \log_{10} (M_{200} , [h^{-1} M_{\odot}]) < 15.5$. The median values are depicted as lines, with the shaded regions representing the $16^{\text{th}}$ to $84^{\text{th}}$ percentiles.

The figure indicates that the central gas distributions of \gizmo{}\_gas differ up to 60\% from \gadget{}\_gas, and the scatter is closer to \gadget{}\_gas as the mass increases. This substantial difference in gas distribution likely plays a crucial role in the U-Net’s ability to predict observables consistent with the hydrodynamical simulations on which it was trained.


\bsp	
\label{lastpage}

\end{document}